\documentclass[a4paper,11pt]{article}
\usepackage{pos}
\usepackage{tikz}
\usetikzlibrary{calc,positioning}
\def\dsqcup{\sqcup\mathchoice{\mkern-7mu}{\mkern-7mu}{\mkern-3.2mu}{\mkern-3.8mu}\sqcup}
\newcommand{\zbar}{\bar{z}}
\newcommand{\zz}{z}
\newcommand{\cC}{\mathcal{C}}
\newcommand{\tildeJ}{\widetilde J}
\def\Li{{\rm Li}}
\title{The Diagrammatic Coaction}
\author[a,b]{Samuel Abreu}
\author[c,d]{Ruth Britto}
\author[e]{Claude Duhr}
\author*[b]{Einan Gardi}
\author[b]{James Matthew}

\affiliation[a]{CERN, Theoretical Physics Department, CH-1211 Geneva 23, Switzerland}
\affiliation[b]{Higgs Centre for Theoretical Physics, 
School of Physics and Astronomy, \\
The University of Edinburgh, Edinburgh EH9 3FD, Scotland, UK}
\affiliation[c]{School of Mathematics, Trinity College, Dublin 2, Ireland}
\affiliation[d]{Hamilton Mathematics Institute, Trinity College, Dublin 2, Ireland}
\affiliation[e]{Bethe Center for Theoretical Physics, Universit\"at Bonn, D-53115, Germany}

\emailAdd{samuel.abreu@cern.ch}
\emailAdd{brittor@tcd.ie}
\emailAdd{cduhr@uni-bonn.de}
\emailAdd{einan.gardi@ed.ac.uk}
\emailAdd{james.matthew01@hotmail.com}

\abstract{The diagrammatic coaction underpins the analytic structure of Feynman integrals, their cuts and the differential equations they admit.
The coaction maps any diagram into a tensor product of its pinches and cuts. These correspond respectively to differential forms defining master integrals, and integration contours which place a subset of the propagators on shell. In a canonical basis these forms and contours are dual to each other. In this talk I review our present understanding of this algebraic structure and its manifestation for dimensionally-regularized Feynman integrals that are expandable to polylogarithms around integer dimensions. Using one- and two-loop integral examples, I will explain the duality between forms and contours, and the correspondence between the \emph{local coaction} acting on the Laurent coefficients in the dimensional regulator and the \emph{global coaction} acting on generalised hypergeometric functions.    }

\FullConference{%
  Loops and Legs in Quantum Field Theory - LL2022,\\
  25-30 April, 2022\\
  Ettal, Germany
}

\begin{document}
\maketitle

\section{Introduction}

In recent years there has been significant progress in understanding the mathematical properties of Feynman integrals. One proposition is that dimensionally-regularized Feynman integrals can be endowed with a coaction, which on the one hand has a purely diagrammatic description, while on the other hand represents an operation on the functions to which the integrals evaluate. 
The most familiar example of such functions are multiple polylogarithms (MPLs)~\cite{GoncharovMixedTate,Goncharov:2005sla}. The latter have a well-established Hopf algebraic structure that includes a coaction~\cite{2002math......8144G,B:MTMZ,2011arXiv1102.1310B}. The coaction on MPLs has been widely used in Feynman integral computations~\cite{Brown:2008um,Anastasiou:2013srw,Panzer:2014caa,Bogner:2014mha,Bogner:2015nda,Ablinger:2014yaa,Duhr:2019tlz}, for example, to understand their analytic continuation  properties, efficiently solve differential equations and algorithmically simplify results. These applications significantly advanced the reach of analytic computation of Feynman integrals over the past decade.  Having witnessed how powerful this algebraic structure is, an interesting question to ask is whether Feynman integrals themselves can be endowed with a coaction, or even whether this is a general property of Feynman integrals. At one loop the answer to this question\footnote{Historically, the search for a combinatorial coaction on Feynman graphs has been an active area of research. Other propositions~\cite{Bloch:2005bh,Kreimer:1997dp,Connes:1998qv,Connes:1999yr,Kreimer:2009jt,Bloch:2010gk,Bloch:2015efx}, however, have not considered dimensional regularization.} is positive
~\cite{Abreu:2017ptx,Abreu:2017enx,Abreu:2017mtm}: dimensionally regularized one-loop Feynman integrals admit a diagrammatic coaction involving pinches and cuts, which maps directly onto the coaction on MPLs.  

The coaction on integrals~\cite{Abreu:2017enx} can be described in general, that is without committing to a particular class of function, as follows. Consider an integral defined by an integrand $\omega$ (a differential form) integrated over a contour $\gamma$. The coaction $\Delta$ acting on this integral takes the form
\begin{align}
\label{int_coaction}
\Delta\left(
\int_{\color[rgb]{0,0.1,1}\gamma}{\color[rgb]{1,0.1,0} \omega }\right)\equiv \sum_{{\color[rgb]{.2,.5,0.2} i}} 
\int_{\color[rgb]{0,0.1,1}\gamma} 
\omega_{\color[rgb]{.2,.5,0.2} i} 
 \otimes 
\int_{\gamma_{\color[rgb]{.2,.5,0.2} i}} {\color[rgb]{1,0.1,0} \omega }\,.
\end{align}
Each left entry of the coaction is characterised by an integrand 
$\omega_{\color[rgb]{.2,.5,0.2} i}$, corresponding to an element of a basis of master integrals, 
$\left\{\int_{\color[rgb]{0,0.1,1}\gamma} 
\omega_{\color[rgb]{.2,.5,0.2} i} \right\}$.
The right entry, in turn, is characterised by the contour 
${\gamma_{\color[rgb]{.2,.5,0.2} i}}$.
The left entries preserve the original contour ${\color[rgb]{0,0.1,1}\gamma}$ while the right ones preserve the original integrand ${\color[rgb]{1,0.1,0} \omega }$.
The coaction on MPLs can be recovered as a special case of this more general formulation.

The aforementioned diagrammatic coaction associates Feynman diagrams  to each of the terms in eq.~(\ref{int_coaction}) as follows.
The integrands $\omega_{\color[rgb]{.2,.5,0.2} i}$ defining each left entry are described by diagrams 
with all or a (non-empty) subset of the propagators of the original integral $\omega$. 
The propagators (edges of the graph) that are absent in a given master integrand are effectively \emph{pinched}, identifying the corresponding vertices. 
The right entry, in turn, is characterised by the contour 
${\gamma_{\color[rgb]{.2,.5,0.2} i}}$ encircling a subset of the propagator poles: each encircled pole places the corresponding particle on-shell, thus \emph{cutting} that edge. An example of this coaction, applied to an off-shell triangle with massless propagators, is given by\footnote{Note that tadpoles are absent in (\ref{eq:triangle}) simply because, being massless, they vanish in dimensional regularization: they do appear if the propagators are massive.}
\begin{align}\label{eq:triangle}
&\Delta\left[\raisebox{-3.7mm}{\includegraphics[keepaspectratio=true, width=1.25cm]{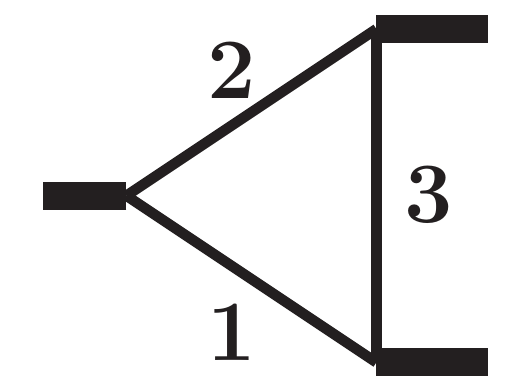}}\right]=
\raisebox{-2.7mm}{\includegraphics[keepaspectratio=true, width=1.2cm]{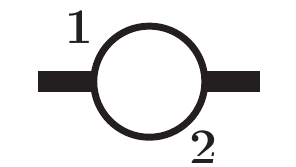}}\otimes\raisebox{-3.7mm}{\includegraphics[keepaspectratio=true, width=1.25cm]{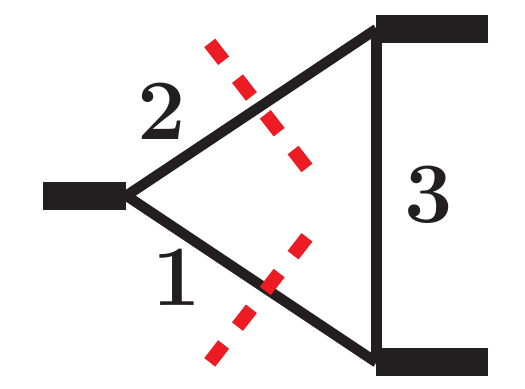}}
+\raisebox{-2.7mm}{\includegraphics[keepaspectratio=true, width=1.2cm]{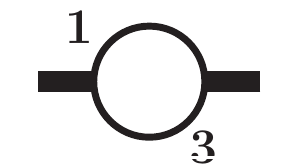}}\otimes\raisebox{-3.7mm}{\includegraphics[keepaspectratio=true, width=1.25cm]{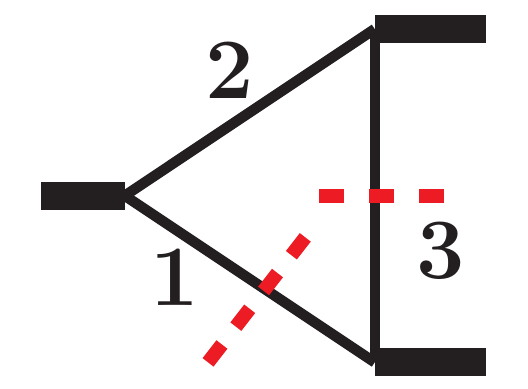}}
+\raisebox{-2.7mm}{\includegraphics[keepaspectratio=true, width=1.2cm]{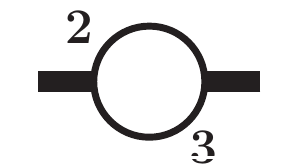}}\otimes\raisebox{-3.7mm}{\includegraphics[keepaspectratio=true, width=1.25cm]{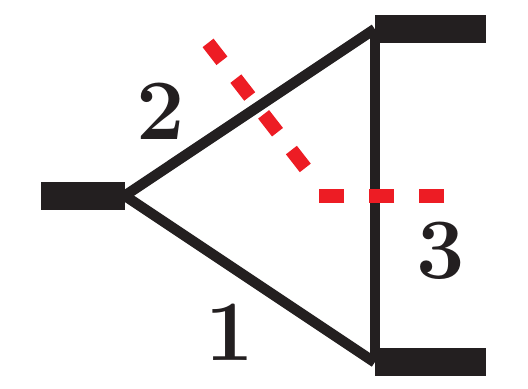}}
+\raisebox{-3.7mm}{\includegraphics[keepaspectratio=true, width=1.25cm]{./diagrams/triPRL.pdf}}\otimes\raisebox{-3.7mm}{\includegraphics[keepaspectratio=true, width=1.25cm]{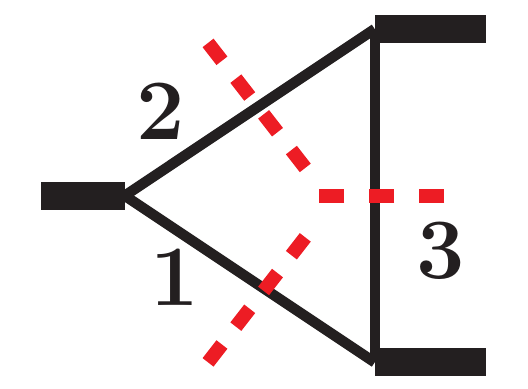}}
\end{align}
where the dashed lines cutting through edges represent cuts. Note that the correspondence between the contour ${\gamma_{\color[rgb]{.2,.5,0.2} i}}$ and the master integrand $\omega_{\color[rgb]{.2,.5,0.2} i}$ is such that only propagators that are cut on the right entry, feature on the corresponding left entry. This is a general condition. 

Refs.~\cite{Abreu:2017enx,Abreu:2017mtm} established the diagrammatic coaction for dimensionally-regularized one-loop Feynman integrals, with any number of edges and any configuration of internal and external masses.  
This line of research continued over the last few years, extending this construction in two ways. First, Refs.~\cite{Abreu:2019xep,Abreu:2019wzk,Brown:2019jng} formulated the coaction on integrals in eq.~(\ref{int_coaction}) directly in terms of hypergeometric functions, thus not requiring an $\epsilon$ expansion. This coaction is sometimes referred to as the \emph{global coaction} and it is consistent with the local coaction acting on MPLs upon expansion in $\epsilon$.
This development benefits from established mathematical techniques to deal with hyperplane arrangements (intersections of hyperplanes) corresponding to the geometry underlying the definition of hypergeometric functions~\cite{gotomatsumoto,Matsumoto1998,Mizera:2017rqa,Mizera:2019gea,Mastrolia:2018uzb}.  
Next, Ref.~\cite{Abreu:2021vhb} took the first steps in generalising the coaction to multi-loop Feynman integrals.

In this talk we will review this topic, starting with a brief discussion of the coaction on MPLs (section~\ref{sec:MPLs}), the properties stemming from the general definition in eq.~(\ref{int_coaction}) (section~\ref{sec:Coaction_prop}), and then describe some central features of the diagrammatic coaction at one loop (section~\ref{sec:1-loop}).
We proceed with one example of the coaction on hypergeometric functions (section~\ref{sec:hyper}) before discussing the generalization of the diagrammatic coaction to two-loop integrals and beyond in section~\ref{sec:two-loop}. We conclude with a short summary and outlook.

\section{Coaction on Multiple Polylogarithms\label{sec:MPLs}}

Multiple Polylogarithms (MPLs) are iterated integrals of the form:
\begin{equation}
\label{Gdef}
G(\vec{a};z)\equiv 
G(\underbrace{a_1,a_2,a_3,...,a_n}_{\text{weight\,\,}n};z)= 
\int_0^z \frac{dt}{t-a_1}G(\underbrace{a_2,a_3...,a_n}_{\text{weight}\,\,n-1};t)\,,
\end{equation}
where the (transcendental) weight corresponds to the number of integrations required to obtain a given MPL starting from a rational function. MPLs admit a shuffle product $G(\vec{a};z)G(\vec{b};z)=\sum_{\vec{c}\in \vec{a}\dsqcup \vec{b} }G(\vec{c};z)$. In addition, as anticipated, they can be endowed with a coaction~\cite{Remiddi:1999ew,GoncharovMixedTate,Brown:2009qja,Brown:2011ik,Duhr:2011zq,Duhr:2012fh}, an operation which maps a given MPL into pairs of simpler (lower weight) MPLs, effectively capturing the algebraic and analytic complexity of these functions.
The coaction on MPLs may be defined by 
\begin{align}
\label{coaction_MPLs}
\Delta(G(\vec c;z)) = \sum_{\vec b\subseteq \vec c} \,\,\,\underbrace{G(\vec b;z)}_{\small \text{weight\,} |\vec b|}\,\,\,\,\otimes\,\,\, \underbrace{G_{\vec b}(\vec c;z)}_{{\small \text{weight\,} |\vec c|-|\vec b|}}\,,
\end{align}
where $\otimes$ stands for a tensor product and $G_{\vec b}(\vec c;z)$ is defined using a modified integration contour, which differs from the original contour of eq.~(\ref{Gdef}) by encircling (precisely) the subset of poles contained in $\vec{b}$.  Note that the left entry $G(\vec b;z)$ has a modified integrand, selecting a subset $\vec{b}$ of the original poles, retaining the original contour, while the corresponding right entry has a modified integration contour encircling these poles, retaining the original integrand defined by the full set of poles $\vec{c}$. 
By breaking a high-weight function into simpler ones, (if applied repeatedly, into logarithms), the coaction allows one to derive functional identities by simple algebra.  
The coaction on MPLs as defined in eq.~(\ref{coaction_MPLs}) provides a template for the more general coaction on integrals (\ref{int_coaction}) and all that follows. 

\section{Coaction on integrals and its properties\label{sec:Coaction_prop}}

The coaction formula of eq.~(\ref{int_coaction}) can be seen as a special case of the definition
\begin{align}
\label{coaction_non-diag}
\Delta\left(
\int_{\color[rgb]{0,0.1,1}\gamma}{\color[rgb]{1,0.1,0} \omega }\right)\equiv \sum_{{\color[rgb]{.5,.2,0.2} i}, {\color[rgb]{.2,.5,0.2} j}} 
{c}_{{\color[rgb]{.5,.2,0.2} i}{\color[rgb]{.2,.5,0.2} j}}\int_{\color[rgb]{0,0.1,1}\gamma} 
\omega_{\color[rgb]{0.5,.2,0.2} i}  \otimes \int_{\gamma_{\color[rgb]{.2,.5,0.2} j}} {\color[rgb]{1,0.1,0} \omega }\,\,.
\end{align}
By rotating the basis of differential forms or the basis of contours, or both, one can recover eq.~(\ref{int_coaction}), corresponding to ${c}_{{\color[rgb]{.5,.2,0.2} i}\!{\color[rgb]{.2,.5,0.2} j}}=\delta_{{\color[rgb]{.5,.2,0.2} i}{\color[rgb]{.2,.5,0.2} j}}$.  The condition ${c}_{{\color[rgb]{.5,.2,0.2} i}{\color[rgb]{.2,.5,0.2} j}}=\delta_{{\color[rgb]{.5,.2,0.2} i}{\color[rgb]{.2,.5,0.2} j}}$ is called the \emph{duality condition}: it identifies a natural pairing between integrands and integration contours. 
The diagonal form of the coaction of eq.~(\ref{int_coaction}) is most convenient and we will see examples of how the duality condition is satisfied in different contexts in the following sections. 

Two important properties of the coaction stem directly from its definition. The fact that the original integration contour is carried by the left entries, dictates how the coaction interacts with a discontinuity operator across branch cuts, namely the discontinuity acts only on the left entries:
\begin{align}
\label{coaction_Disc} 
\Delta\left({\rm Disc}\left[
\int_{\color[rgb]{0,0.1,1}\gamma}{\color[rgb]{1,0.1,0} \omega }\right]\right)= \sum_i 
{\rm Disc}\left[\int_{\color[rgb]{0,0.1,1}\gamma} 
\omega_i\right]  \otimes \int_{\gamma_i} {\color[rgb]{1,0.1,0} \omega }\,.
\end{align}
Similarly, the fact that the original integrand is in the right entries, dictates the interaction of the coaction with differentiation, namely derivatives act only on the right entries: 
\begin{align}
\label{coaction_Diff} 
\Delta\left({\rm \partial}\left[
\int_{\color[rgb]{0,0.1,1}\gamma}{\color[rgb]{1,0.1,0} \omega }\right]\right)= \sum_i 
\int_{\color[rgb]{0,0.1,1}\gamma} 
\omega_i\otimes{\rm \partial}\left[\int_{\gamma_i} {\color[rgb]{1,0.1,0} \omega }\right]  \,.
\end{align}
Eq.~(\ref{coaction_Disc}) is therefore key to performing analytic continuation, while eq.~(\ref{coaction_Diff}) to deriving differential equations. Both these operations are essential in the context of Feynman integrals, which we consider next.

\section{The diagrammatic coaction at one loop\label{sec:1-loop}}

One-loop integrals, considered in dimensional regularization, and with any internal and external configuration of masses, present  a rich class of examples. In this section we briefly describe the form the coaction~(\ref{int_coaction}) takes in these cases following Refs.~\cite{Abreu:2017enx,Abreu:2017mtm}.
A particular advantage of one-loop integrals, is that it is easy to specify a pure basis at the outset. Specifically, it is sufficient to consider scalar integrals with single-power propagators, as this set of integrals forms a basis for any one-loop integral. Furthermore, considering the case of dimensional regularization around an even number of dimensions
we may consider one-loop integrals with an even number of propagators $n_{\text{even}}$ in $D=n_{\text{even}}-2\epsilon$ dimensions, and ones with an odd number of propagators $n_\text{odd}$ in $D=(n_{\text{odd}}+1)-2\epsilon$ dimensions. That is, for any $n$ we fix the number of dimensions as $D=2\lceil\frac{n}{2}\rceil-2\epsilon$. Thus, tadpoles and bubbles are considered in $2-2\epsilon$ dimensions, triangles and boxes in $4-2\epsilon$ dimensions, etc.
Dimensional-shift relations allow one to obtain expressions in other even dimensions. Importantly, all these basis integrals, which we denote by $\tildeJ_n$, are \emph{pure functions} \cite{Arkani-Hamed:2010pyv}
of uniform transcendental weight (assuming we assign $\epsilon$ weight $-1$), up to an overall multiplicative kinematic factor. 

We have already seen one example, that of the off-shell triangle in eq.~(\ref{eq:triangle}) above.\footnote{Note that all diagrams appearing there represent elements in the aforementioned pure basis: the triangles are in defined in $4-2\epsilon$ dimensions , while the bubbles are defined in $2-2\epsilon$ dimensions.} To understand how the diagrammatic coaction maps (exactly, but in a highly non-trivial way) onto a coaction on MPLs, let us consider the coaction on the leading-order term in the $\epsilon$ expansion of eq.~(\ref{eq:triangle}). At leading order in $\epsilon$ the triangle evaluates to the Bloch-Wigner single-valued dilogarithm, see e.g. Ref.~\cite{Chavez:2012kn}:
\begin{equation}\label{tp1p2p3Ep0}
	\mathcal{T}(\zz,\zbar)=-2\Li_2(\zz)+2\Li_2(\zbar)-\ln(\zz\zbar)
	\ln\left(\frac{1-\zz}{1-\zbar}\right)\,,
\end{equation}
where 
%\begin{equation}
%\label{eq:ZZbarDef}
$\zz\,\zbar = \frac{p_2^2}{p_1^2} {\rm~~and~~} (1-\zz)\,(1-\zbar) = \frac{p_3^2}{p_1^2}$\,.
%\end{equation}
The coaction on $\mathcal{T}(\zz,\zbar)$ reads:
\begin{align}
\label{eq:trianglecoproduct}
	\Delta\left[\mathcal{T}(\zz,\zbar)\right]
%	=\,&\mathcal{T}(\zz,\zbar)\otimes1 + 1\otimes\mathcal{T}(\zz,\zbar) +
%	\ln(\zz\bar\zz)\otimes \ln\frac{1-\bar\zz}{1-\zz}\\
%\nonumber	\,&+ \ln[(1-\zz)(1-\bar\zz)] \otimes \ln\frac{\zz}{\bar\zz}\nonumber \\
%\nonumber	
=\,& \ln\left(-p_2^2\right)\otimes
	\ln\frac{1-\bar\zz}{1-\zz} + \ln\left(-p_3^2\right)\otimes \ln\frac{\zz} {\bar\zz} + \ln (-p_1^2) \otimes  \ln\frac{\zbar(1-\zz)}{\zz(1-\zbar)}
\nonumber\\&
+\mathcal{T}(\zz,\zbar)\otimes1 + 1\otimes\mathcal{T}(\zz,\zbar) \,,
\end{align}
where the logarithmic terms in the left entries of the first line have been re-expressed  in terms of Mandelstam invariants,  manifesting the \emph{first-entry condition}~\cite{Gaiotto:2011dt,Abreu:2014cla,Abreu:2017mtm}: these terms capture the discontinuity in the three channels according to eq.~(\ref{coaction_Disc}).
It is clear that these logarithmic terms arise from the terms containing bubble integrals in the left entry in eq.~(\ref{eq:triangle}) which indeed have a unitarity cut in the respective channel. 
It is also clear that the penultimate term in eq.~(\ref{eq:trianglecoproduct}), with $\mathcal{T}(\zz,\zbar)$ in the left entry, represents the triangle of eq.~(\ref{eq:triangle}). It is harder to understand the origin of the remaining term, $1\otimes\mathcal{T}(\zz,\zbar)$. Its origin becomes clear~\cite{Abreu:2017mtm}, however, considering  that eq.~(\ref{eq:triangle}) holds in dimensional regularization, and bubble integrals also have a divergent $1/\epsilon$ contributions (these are bubbles in $2-2\epsilon$ dimensions, which are infrared divergent). These $1/\epsilon$ poles multiply the two-propagator cuts of the triangle shown in the right entries in eq.~(\ref{eq:triangle}), and a finite ${\cal O}(\epsilon^0)$ contribution to the coaction arises then from the ${\cal O}(\epsilon^1)$ term in the expansion of these cuts. Remarkably these add up to reproduce the term $1\otimes\mathcal{T}(\zz,\zbar)$. Moreover, the linear relation underlying this cancellation generalises to all orders in $\epsilon$ and can be understood as a diagrammatic relation between cut integrals:\footnote{This relation holds up to $i\pi$ terms, which are irrelevant on the right entry.}
\begin{figure}[h!]
\begin{center}
     \includegraphics[width=.85\textwidth]{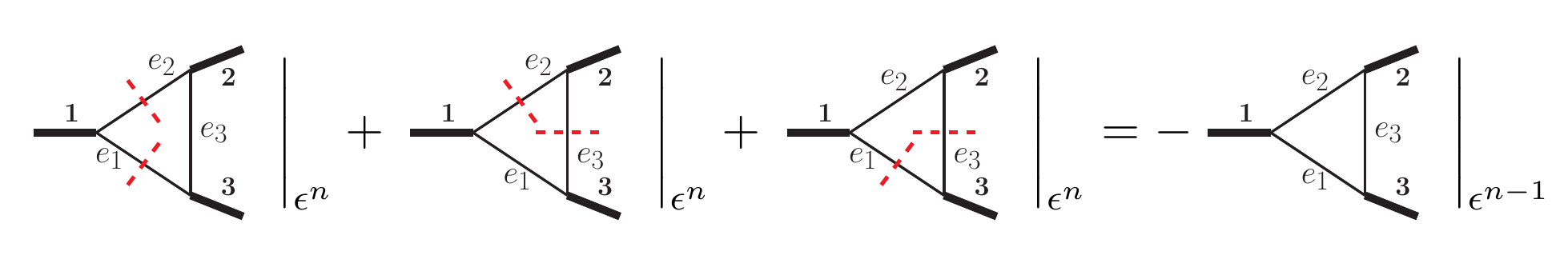}
     \caption{The relation between cuts in the case of the off-shell triangle with massless propagators.}
     \label{fig:triangle_pole_cancellation}
\end{center}
\end{figure}

The triangle example gave us an opportunity to appreciate the non-trivial nature of the diagrammatic coaction map, relating the purely diagrammatic statement of eq.~(\ref{eq:triangle}), to a statement regarding the polylogarithmic function the corresponding (cut) integrals evaluate to in dimensional regularization.  Refs.~\cite{Abreu:2017enx,Abreu:2017mtm} demonstrated that such a precise mapping exists for any one-loop integral. In particular, a similar relation to Figure~\ref{fig:triangle_pole_cancellation} applies to any one-loop integral $\tildeJ_n$ with $n$ propagators. It reads:  
\begin{equation}\label{eq:PCI}
\sum_{i\in [n]} \cC_{i}\tildeJ_n + \sum_{\substack{i,j\in [n],\, i<j}}\cC_{ij}\tildeJ_n = -\epsilon\,\tildeJ_n\mod i\pi\,,
\end{equation}
where  $\cC_{i}$ represents a cut operator acting on propagator $i$ and $[n]=\{1,2,\ldots,n\}$. This implies that any integral $\tildeJ_n$ can be recovered (up to $i\pi$ terms) through a sum of its single and double propagator cuts.
This identity explains why the uncut integral itself is not ever required on the right entry of the coaction. As explained in Section 2 in Ref.~\cite{Abreu:2017mtm}, eq.~(\ref{eq:PCI}) is a specific manifestation of the one-loop \emph{homology relation}~\cite{Fotiadi1,Abreu:2017ptx} generalised to dimensional regularization, which is central to the structure of the diagrammatic coaction. 

Let us turn then to the formulation of the diagrammatic coaction at one loop, which we view as a special case of the coaction on integrals in eq.~(\ref{int_coaction}).  Consider the integrand of a generic one-loop graph $G$ with $n$ propagators, which we denote by $\color[rgb]{1.0,0.1,0.1}\omega_G$, such that 
\[
\tildeJ_n = \int_{{\color[rgb]{0.1,0.1,1.0}\Gamma_\emptyset}} {\color[rgb]{1.0,0.1,0.1}\omega_G} \,,
\]
where we denote the usual loop-momentum integration contour ${{\color[rgb]{0.1,0.1,1.0}\Gamma_\emptyset}}$, corresponding to the unrestricted integration over all momentum components. The coaction of this generic one-loop integral takes the form:
\begin{align}
\label{Diagr_coaction_one_loop}
\Delta\left(\int_{{\color[rgb]{0.1,0.1,1.0}\Gamma_\emptyset}}{\color[rgb]{1.0,0.1,0.1}\omega_G}\right)=
	\sum_{C\in M_G}
	%{i=1}^{\displaystyle{\kappa_{G_C}}}
	\int_{{\color[rgb]{0.1,0.1,1.0}\Gamma_\emptyset}}\omega_{G_C}\otimes
	\int_{\gamma_C}{\color[rgb]{1.0,0.1,0.1}\omega_G}\qquad\text{with}\qquad \gamma_C=\left\{
                \begin{array}{ll}
                    \Gamma_{C\infty} & \rm{for} \,\,|C|\,\, {\rm odd} \\
                    \Gamma_C         & \rm{for} \,\, |C|\,\, {\rm even} 
                \end{array}
              \right. \,,
\end{align}
where $C$ is a non-empty subset of propagators in $G$ and $\omega_{G_C}$ in the left entry is the master integrand corresponding to the graph $G_C$, which is obtained from the original graph $G$ by pinching all the propagators which are not in the set $C$. For later convenience we also introduce the notation $M_G$, 
denoting the basis of master topologies of $G$, which at one loop simply consists of $G$ itself and all its pinches (those that vanish in dimensional regularization may be excluded).  
Finally, the master contour $\gamma_C$ corresponding to $\omega_{G_C}$, defined in eq.~(\ref{Diagr_coaction_one_loop}),
satisfies the duality condition $\int_{\gamma_C}\omega_{G_C} =1+{\cal O}(\epsilon)$, while 
for other contours $\gamma_{C^\prime}$ in the basis ($C^\prime\in M_G$), $\int_{\gamma_{C^{\prime}}}\omega_{G_C} ={\cal O}(\epsilon)$.
Writing  $\gamma_C$ in eq.~(\ref{Diagr_coaction_one_loop}) we used the notation $\Gamma_C$ for the contour that encircles (precisely) the poles of the propagators in the set~$C$, thus effectively cutting the corresponding edges of the graph. Similarly $\Gamma_{C\infty}$ encircles the same set of propagator poles, plus the singularity at infinite loop momentum. It was further shown~\cite{Abreu:2017mtm} that the aforementioned homology relations allow one to express any integral over the contour $\Gamma_{C\infty}$ (up to $i\pi$ terms) as the following linear combination of contours that do not involve the pole at infinity,  
\begin{equation}
\label{gamma_C}
\gamma_C\equiv \Gamma_C+a_C\sum_{e\notin  C}\Gamma_{Ce}\,,
\qquad\text{with}\qquad
a_C=\left\{
\begin{array}{ll}
\frac12 &\text{if }|C|\text{ is odd}\\
0 &\text{if }|C|\text{ is even}\,.
\end{array}
\right.
\end{equation}
The conclusion is that with our choice of basis integrands $\omega_{G_C}$ for the left entries,
the right entries are simply the cuts of the original integral where, for  even~$|C|$, \emph{only} the propagators in $C$ are cut, while for odd $|C|$ it is this cut plus
a half times the sum of cuts in which one additional propagator that is not in $C$ is cut as well. In every term, all the propagators that feature on the left entry, must be cut on the corresponding right entry.
\begin{figure}[h!]
\begin{center}
     \includegraphics[width=1\textwidth]{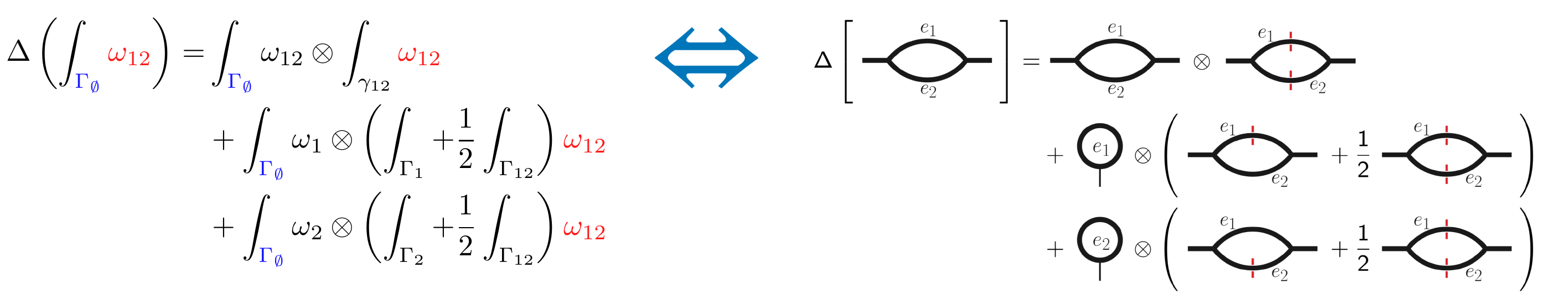}
     \caption{The diagrammatic coaction of the bubble integral with massive propagators.}
     \label{fig:two-massBubbleCoation}
\end{center}
\end{figure}
%\begin{align}
%\begin{split}
%\Delta\left(\int_{{\color[rgb]{0.1,0.1,1.0}\Gamma_\emptyset}}{\color[rgb]{1.0,0.1,0.1}\omega_{12}}\right)=&
%	\int_{{\color[rgb]{0.1,0.1,1.0}\Gamma_\emptyset}}\omega_{12}\otimes
%	\int_{\gamma_{12}}{\color[rgb]{1.0,0.1,0.1}\omega_{12}}
%\\
%  &+\int_{{\color[rgb]{0.1,0.1,1.0}\Gamma_\emptyset}}\omega_{1}\otimes
%\left( \int_{\Gamma_{1}}+\frac12 \int_{\Gamma_{12}}\right){\color[rgb]{1.0,0.1,0.1}\omega_{12}}
%\\  &+\int_{{\color[rgb]{0.1,0.1,1.0}\Gamma_\emptyset}}\omega_{2}\otimes\left( \int_{\Gamma_{2}}+\frac12 \int_{\Gamma_{12}}\right){\color[rgb]{1.0,0.1,0.1}\omega_{12}}
%\end{split}
% \end{align}
As a final example, consider the bubble integral in Figure~\ref{fig:two-massBubbleCoation}, illustrating both even and odd $|C|$.

\begin{figure}[h!]
\begin{center}
     \includegraphics[width=.8\textwidth]{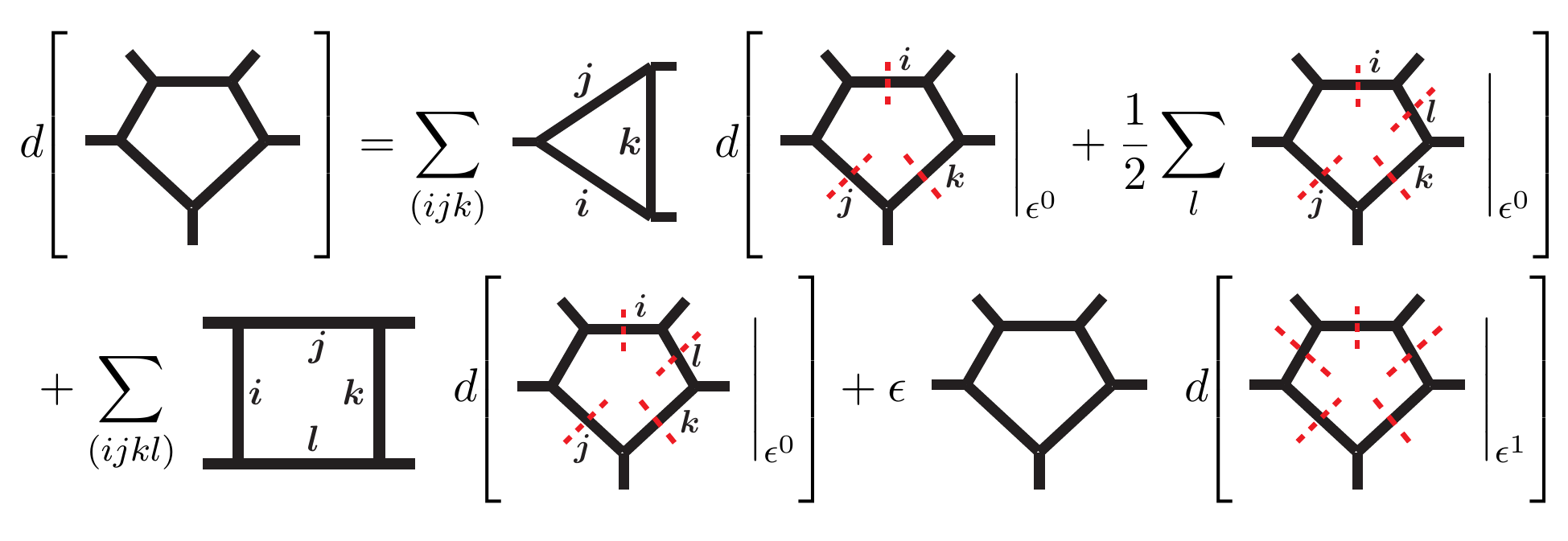}
     \caption{The differential equation for a pentagon integral, as derived from the diagrammatic coaction.}
     \label{fig:Pentagon_DE}
\end{center}
\end{figure}

One application of the diagrammatic coaction discussed in Ref.~\cite{Abreu:2017mtm} is the derivation of differential equations. The key observation is that because differentiation only acts on the right entries, as in eq.~(\ref{coaction_Diff}), when using a pure basis (as we do) the derivative of a given Feynman integral can be determined in full from terms in the coaction in which the right entry has weight 1. After differentiation, such terms yield rational functions which simply multiply the respective master integral in the left entry. This means that the coefficients of the differential equation are directly determined by derivatives of certain (maximal, next-to-maximal and next-to-next-to-maximal) cuts, computed at the relevant order in the $\epsilon$ expansion to yield weight 1 contributions. As an example, we show in Figure~\ref{fig:Pentagon_DE} the case of a generic pentagon. The result for a generic one-loop integral can be found in Ref.~\cite{Abreu:2017mtm}.

This concludes our brief exposition of the one-loop diagrammatic coaction. We have seen that the duality condition is realised in a rather straightforward manner, and the dual contours can be expressed in terms of ordinary cut integrals with at least those propagators present on the left entry being put on shell. Specifying the precise linear combination of cuts is based on the homology relations. Finally, we have seen how the diagrammatic coaction encodes the differential equations.
 
\section{The coaction on hypergeometric functions\label{sec:hyper}}

So far our interpretation of the coaction formula in terms of functions was based exclusively on MPLs.
Therefore, the application to Feynman integrals was based on applying the coaction to the functions appearing order by order in the Laurent expansion in $\epsilon$. This coaction is sometimes referred to as the \emph{local coaction}~\cite{Brown:2019jng}.
As emphasised already in Ref.~\cite{Abreu:2017ptx}, eq.~(\ref{int_coaction}) itself is more general. In particular, it may be applied directly to the hypergeometric functions one obtains in dimensional regularization, without considering the expansion in $\epsilon$. This is called the \emph{global coaction}.
With this motivation, Refs.~\cite{Abreu:2019xep,Abreu:2019wzk,Brown:2019jng} investigated the application of the coaction directly on hypergeometric functions. Importantly, the global and local coactions must be consistent. Specifically, when the Laurent coefficients are MPLs, it is expected that by expanding both left and right entries in the coaction on hypergeometric functions one would reproduce the coaction on MPLs. 
This correspondence was proven for certain classes of functions~\cite{Brown:2019jng}.

Let us make some comments regarding the formulation of the global coaction. The mathematical framework of twisted (co-)homology theory allows one to form a basis of differential forms~$\omega_j$ (generating the cohomology group) and a corresponding basis of integration contours~$\gamma_i$ (generating the homology group), such that their pairing, $\int_{\gamma_i}\omega_j$, defines a class of hypergeometric-type integrals~\cite{AomotoKita}. With these bases in place one can define a coaction on any function in this space, where all the entries in the coaction are expressed in terms of the same class of function. Specifically, to express the integrand one defines 
%\begin{align}
%\omega_j= {\color[rgb]{0.1,0.1,1.0} \Psi({\bf u})}{\color[rgb]{1.0,0.1,0.1}\varphi_{n_1\ldots n_K}({\bf u})} ,\qquad 
%{\color[rgb]{0.1,0.1,1.0} \Psi({\bf u}) = \prod_{I} P_I({\bf u})^{a_I \epsilon}}\quad \textrm{and} \quad
%{\color[rgb]{1.0,0.1,0.1}\varphi_{n_1\ldots n_K}({\bf u}) = d{\bf u} \prod_{I=1}^K P_I({\bf u})^{n_I}}
%\end{align}
\begin{align}
\label{integrand_factor}
\omega_j=  \Phi({\bf u}) \varphi_{n_1\ldots n_K ({\bf u})} \quad \text{with}\quad
\Phi({\bf u}) = \prod_{I=1}^{K} P_I({\bf u})^{a_I \epsilon} \quad \textrm{and} \quad
\varphi_{n_1\ldots n_K}({\bf u}) = d{\bf u} \prod_{I=1}^K P_I({\bf u})^{n_I}
\end{align}
in terms of polynomials $P_I({\bf u})$, where ${\bf u}$ is a vector of integration variables.  Assuming that $\epsilon$ is real while the indices~$n_I$ are integers, multivaluedness of the integrand $\omega_j$ is controlled exclusively by $\ln  \Phi({\bf u})$, the so-called twist.
The integration contours $\gamma_i$ are then defined between the zeros of polynomials $P_I({\bf u})$.
Geometrically, these correspond to hypersurfaces, often hyperplanes, whose arrangement (and intersections) fully characterise the class of integrals. Methods from intersection theory~\cite{gotomatsumoto,Matsumoto1998,Mizera:2017rqa,Mizera:2019gea,Mastrolia:2018uzb}
then become handy in computing the coaction coefficients ${c}_{{\color[rgb]{.5,.2,0.2} i}{\color[rgb]{.2,.5,0.2} j}}$ of eq.~(\ref{coaction_non-diag}). This also allows one to pick bases of forms and contours that satisfy the duality condition, bringing the coaction to the diagonal form of eq.~(\ref{int_coaction}).

Several classes  of generalised hypergeometric functions have been analysed in Ref.~\cite{Abreu:2019wzk} along similar lines,  including ${}_{p+1}F_p$, Lauricella $F_D$ and the four Appell functions. Here we restrict ourselves to quoting the results of one simple example, namely Gauss hypergeometric function.  We start with the integral representation of the latter, involving a single integration variable $u$,
\begin{align}
\label{eq:2f1Simp}
{}_2F_1(\alpha,\beta;\gamma;x) = \frac{\Gamma(\gamma)}{\Gamma(\alpha)\Gamma(\gamma-\alpha)}\int_0^1 du u^{\alpha-1} (1-u)^{\gamma-\alpha-1} (1-ux)^{-\beta}\,,
\end{align}
which leads, according to eq.~(\ref{integrand_factor}), to the following identification 
\begin{align}
\Phi(u) = u^{a\epsilon} (1-u)^{(c-a)\epsilon} (1-xu)^{-b \epsilon}\,, \quad
\varphi_{n_{\alpha}n_{\beta}n_{\gamma}} (u)= u^{n_\alpha-1} (1-u)^{n_{\gamma}-n_{\alpha}-1} (1-xu)^{-n_{\beta}}\,du\,,
\end{align}
where $\alpha=n_\alpha+a\epsilon$, $\beta=n_\beta+b\epsilon$ and
$\gamma=n_\gamma+c\epsilon$. 
For integer indices $n_\alpha$, $n_\beta$ and $n_\gamma$, eq.~\eqref{eq:2f1Simp} defines a meromorphic 
function of $\epsilon$, with Laurent coefficients that are linear combinations of MPLs with rational coefficients. The homology and cohomology groups associated with this function are two-dimensional. We choose a basis of contours going between the zeros of the polynomials in $\Phi(u)$,
\begin{equation}
\label{Gamma_basis_2F1}
	\gamma_1=[0,1]\,,\qquad
	\gamma_2=[0,1/x]\,,
\end{equation}
and a dual basis for the integrands, 
\begin{align}
\label{omega_2F1}
\begin{split}
	&\omega_1=(c-a)\epsilon\,\Phi(u)\varphi_{101}(u) = (c-a)\epsilon\,u^{a\epsilon} (1-u)^{-1+(c-a)\epsilon}
	(1-xu)^{-b\epsilon}du\,,\\
	&\omega_2= -b\epsilon x\,\Phi(u)\varphi_{112}(u) = -b\epsilon x\,u^{a\epsilon} (1-u)^{(c-a)\epsilon}
	(1-xu)^{-1-b\epsilon}du\,.
\end{split}
\end{align}
With this choice $\int_{\gamma_{\color[rgb]{.5,.2,0.2} i}}\omega_{\color[rgb]{.2,.5,0.2} j} = \delta_{{\color[rgb]{.5,.2,0.2} i}{\color[rgb]{.2,.5,0.2} j}}+{\cal O}(\epsilon)$ and then
 ${c}_{{\color[rgb]{.5,.2,0.2} i}{\color[rgb]{.2,.5,0.2} j}}=\delta_{{\color[rgb]{.5,.2,0.2} i}{\color[rgb]{.2,.5,0.2} j}}$ in
eq.~(\ref{coaction_non-diag})  leading to a diagonal coaction as in eq.~(\ref{int_coaction}), given by~\cite{Abreu:2019wzk}
\begin{align}
\nonumber\Delta\Big({}_2F_1(\alpha,\beta;\gamma;x)\Big) &=
{}_2F_1(1+a\epsilon,b\epsilon;1+c\epsilon;x) \otimes {}_2F_1(\alpha,\beta;\gamma;x) \\
\label{eq:coaction2F1}&- \frac{b\epsilon}{1+c\epsilon}\,
{}_2F_1(1+a\epsilon,1+b\epsilon;2+c\epsilon;x) \\
\nonumber& ~~\otimes \frac{\Gamma(1-\beta)\Gamma(\gamma)}{\Gamma(1-\beta+\alpha)\Gamma(\gamma-\alpha)}
x^{1-\alpha}{}_2F_1\left(\alpha,1+\alpha-\gamma;1-\beta+\alpha;\frac{1}{x}\right)\,.
\end{align}
Naturally, the particular form of the coaction in eq.~\eqref{eq:coaction2F1}
depends on the choice of bases made for the contours and integrands. Other choices would lead to equivalent formulae, related
to eq.~\eqref{eq:coaction2F1} through contiguous and analytic
continuation relations. 
As anticipated, for integer $n_{\alpha}, n_\beta, n_\gamma$, eq.~\eqref{eq:coaction2F1} 
can be expanded into a Laurent series involving only MPLs. Its expansion is consistent with computing the local coaction on MPLs at each order in the expansion. This was conjectured in Ref.~\cite{Abreu:2019wzk}, and proven in Ref.~\cite{Brown:2019jng}.
Finally, we stress a simple but important lesson from the study of the coaction on hypergeometric functions, namely that all the elements of the coaction, both left and right entries, may be expressed in terms of the same class of function. This property becomes useful when studying the coaction on multiloop Feynman integrals, as discussed below.

\section{Constructing the coaction on two-loop integral families\label{sec:two-loop}}

The availability of the global coaction on hypergeometric functions opens the way to extending the diagrammatic coaction to Feynman integrals beyond one loop. It is clear at the outset that this generalization is highly non-trivial, as one-loop integrals are special in several ways. 
First, all one-loop Feynman integrals evaluate to MPLs, making the application of the local coaction straightforward in principle.  In contrast, starting at two loop, the Laurent expansion of Feynman integrals may include elliptic polylogarithms (and more complicated iterated integrals) as well. Work on the generalization of the diagrammatic coaction to the elliptic case is in progress, and is beyond the scope of this talk, but we mention that the global coaction provides a good starting point.
Second, as described in section~\ref{sec:1-loop}, all one-loop Feynman integrals have a natural basis  of master integrals (and a corresponding basis of cuts). In particular, this basis consists of a single integral for a given (sub-)set of propagators: specifying the propagators uniquely identifies the master integral. Similarly, specifying the subset of propagators that is being cut, uniquely identifies an integration contour. 
Starting at two loops an a priori basis is not available: the basis needs to be determined for each topology, e.g. by studying the solutions of the corresponding set of integration-by-parts identities. Furthermore, in many cases, multi-loop integrals have multiple master integrals corresponding to the same set of propagators, and a basis requires raising propagators to higher powers or including numerators that depend on the loop momenta through irreducible scalar products. 
This presents a significant challenge to the generalization of the diagrammatic coaction even in the purely polylogarithmic case. Here we follow Ref.~\cite{Abreu:2021vhb} and demonstrate how this issue is addressed. 

The approach taken by Ref.~\cite{Abreu:2021vhb} is to construct the  diagrammatic coaction on a case-by-case basis by taking full advantage of the availability of explicit coaction formulae for a wide class of generalised hypergeometric-type functions. 
One begins by considering a particular integral topology, 
establishes a basis of master integrals and evaluates these in terms of hypergeometric functions. 
Next one computes the cuts of a given integral in this space, whose diagrammatic coaction is of interest. 
Based on what we have learnt about the coaction of hypergeometric functions, the expectation is that all master integrals, as well as all cuts should be expressible in terms of the same class of function.
 This is indeed what one finds in every example considered.
The next step is therefore clear: one evaluates the global coaction of the Feynman integral considered using the known coaction formulae of hypergeometric-type functions~\cite{Abreu:2019wzk}, and then expresses the left entries in terms of the master integrals and the right entries in terms of the cuts. Once this is achieved, the result has an immediate diagrammatic interpretation. This step may be non-trivial, as it requires using both contiguous relations (integration-by-parts identities) as well as analytic continuation identities to bring the left and right entries of the coaction into a form where the respective basis elements are identified. Nevertheless, in practice this was achieved in every case considered.

To demonstrate the application of the diagrammatic coaction to two-loop Feynman integrals, let us focus first on a simple example, namely the sunset integral with one massive propagator. This topology is defined by 
\begin{equation}\label{eq:ssetFamily}
S(\nu_1,\nu_2,\nu_3,\nu_4,\nu_5;D;p^2,m^2)=\left(\frac{e^{\gamma_E\epsilon}}{i\pi^{D/2}}\right)^2
\int d^Dk\,d^Dl
\frac{[(k+l)^2]^{-\nu_4}[(l+p)^2]^{-\nu_5}}{[k^2]^{\nu_1}[l^2]^{\nu_2}[(k+l+p)^2-m^2]^{\nu_3}},
\end{equation}
for integer $\nu_i$ with $\nu_4,\nu_5\leq0$ and for $D=n-2\epsilon$, with $n$ even.
This space is known to be two-dimensional, and we choose as basis elements
\begin{subequations}
\begin{align}
\label{S1}
S^{\color[rgb]{1,0.1,0}  (1)}(p^2,m^2)&=\epsilon^2\left({p^2}-{m^2}\right)
S(1,1,1,0,0;2-2\epsilon;p^2,m^2)\\
\nonumber&=(m^2)^{-2\epsilon}\left(1-\frac{p^2}{m^2}\right)e^{2\gamma_E\epsilon}
\Gamma(1+2\epsilon)\Gamma(1-\epsilon)\Gamma(1+\epsilon)
\,{}_2F_1\left(1+2\epsilon,1+\epsilon;1-\epsilon;\frac{p^2}{m^2}\right)
\\
\begin{split}
\label{S2}
S^{\color[rgb]{0.1,0.1,1}  (2)}(p^2,m^2)&\,=\,-\epsilon^2S(1,1,1,-1,0;2-2\epsilon;p^2,m^2)\\
&\,=\,(m^2)^{-2\epsilon}e^{2\gamma_E\epsilon}\Gamma(1+2\epsilon)
\Gamma(1-\epsilon)\Gamma(1+\epsilon)
\,{}_2F_1\left(2\epsilon,\epsilon;1-\epsilon;\frac{p^2}{m^2}\right)\,,
\end{split}
\end{align}
\end{subequations}
which are both pure functions. The result is expressed in terms Gauss hypergeometric functions, whose coaction we know from eq.~(\ref{eq:coaction2F1}) above. In order to be able to identify the right entries in the coaction in terms of cut diagrams, let us first evaluate the cuts.

The maximal cut of the first master integral, eq.~(\ref{S1}), is
\begin{equation}\label{OneMassSunsetMaxCutwoNormalization}
	\mathcal{C}_{1,2,3}S^{(1)}\sim\,
	\mathcal{C}_1\int\frac{d^{2-2\epsilon}k}{i\pi^{1-\epsilon}}
	\frac{1}{k^2}\,\,\left(
	\mathcal{C}_{2,3}\int\frac{d^{2-2\epsilon}l}{i\pi^{1-\epsilon}}
	\frac{1}{l^2}\frac{1}{(k+l+p)^2-m^2}\right)\,,
\end{equation}
where we use the notation $\mathcal{C}_i$ (as in eq.~(\ref{eq:PCI})) for a cutting operation of propagator $i$; rather than equality we used here the symbol $\sim$ because we do not keep track here
of overall normalisation factors  -- we refer the reader to appendix A of Ref.~\cite{Abreu:2021vhb} for details.
The expression in parentheses in eq.~(\ref{OneMassSunsetMaxCutwoNormalization}) is readily identified as
the maximal cut of a one-loop bubble integral
with a single massive propagator and a massive external leg
of mass $(k+p)^2$. Using the result for this one-loop cut integral~\cite{Abreu:2017ptx} we get
\begin{equation}\label{eq:tempCut23}
	\mathcal{C}_{1,2,3}S^{(1)}\sim\,
	\mathcal{C}_1\int\frac{d^{2-2\epsilon}k}{i\pi^{1-\epsilon}}
	\frac{1}{k^2}\,\,\Big[(k+p)^2-m^2\Big]^{-1-2\epsilon}
	\Big[(k+p)^2\Big]^{\epsilon}\,.
\end{equation}
The integrand of the remaining integral is similar to the integrand
of a one-loop one-mass bubble integral. We can thus use one-loop techniques~\cite{Abreu:2014cla,Abreu:2015zaa,Abreu:2017ptx} to compute its cut,  getting
\begin{equation}\label{eq:tempCut123}
	\mathcal{C}_{1,2,3}S^{(1)}\sim
	\int dk_0\,k_0^{-1-2\epsilon}
	\left(p^2-m^2+2\sqrt{p^2}k_0\right)^{-1-2\epsilon}
	\left(p^2+2\sqrt{p^2}k_0\right)^{2\epsilon}\,.
\end{equation}
At this stage we encounter a major difference between one-loop 
cuts and multi-loop ones. After having imposed all cut
conditions, we have not fully localised the integrand. Instead, we have a one-dimensional integral left to perform.
The integration region over $k_0$ has not been specified in eq.~(\ref{eq:tempCut123}) because it is not determined by the cut conditions.  However, knowing that the space of master integrals is two dimensional, we also expect two independent cuts.  
We also recognise that the integrand in (\ref{eq:tempCut123}) is compatible with the integral representation of the Gauss hypergeometric function (\ref{eq:2f1Simp}), which has a two-dimensional homology space. With this in mind we may choose a basis of contours, e.g.
\begin{equation}
\label{Gamma12_one-masssunset_def}
	\Gamma^{(1)}_{1,2,3}:\,\,k_0\in\left[-\frac{\sqrt{p^2}}{2},0\right]\,,
	\qquad \quad
	\Gamma^{(2)}_{1,2,3}:\,\,k_0\in\left[\frac{m^2-p^2}{2\sqrt{p^2}},0\right]\,.
\end{equation}
With these we may readily evaluate two independent maximal cuts of each of the two master integrals.
However, in order to bring the coaction into a simple, diagonal form we require the duality condition 
\begin{equation}
\int_{\gamma^{(j)}_{1,2,3}}\omega^{(i)} = \delta_{ij}+{\cal O}(\epsilon)\,.
\end{equation}
to be satisfied, where $\omega^{(i)}$ correspond to the two master integrands of eq.~(\ref{S1}) and (\ref{S2}) respectively. The duality condition is satisfied for
$\gamma^{(1)}_{1,2,3}$ and $\gamma^{(2)}_{1,2,3}$ that are related to those of eq.~(\ref{Gamma12_one-masssunset_def}) by
\begin{equation}\label{eq:homBasisSunset}
	\gamma^{(1)}_{1,2,3}=\frac{1}{4\epsilon}\Gamma^{(2)}_{1,2,3}\,,\qquad
	\gamma^{(2)}_{1,2,3}=\frac{1}{2\epsilon}
	\left(\Gamma^{(1)}_{1,2,3}-\frac{1}{2}\Gamma^{(2)}_{1,2,3}\right)\,,
\end{equation}
which yields the coaction
\begin{align}
\label{OneMassSunset_coaction}
\Delta \int_{\Gamma_{\emptyset}} \omega^{(i)} =    
\int_{\Gamma_{\emptyset}} \omega^{\color[rgb]{1.0,0.1,0.1}(1)}
\otimes
\int_{\gamma_{1,2,3}^{{\color[rgb]{1.0,0.1,0.1}(1)}}}\omega^{(i)} 
\,\,+\,\,
\int_{\Gamma_{\emptyset}} \omega^{\color[rgb]{.1,0.1,1.0}(2)}
\otimes
\int_{\gamma_{1,2,3}^{{\color[rgb]{0.1,0.1,1.0}(2)}}}\omega^{(i)} \,.
\end{align}
Here the left entries correspond to the two master integrals of eq.~(\ref{S1}) and~(\ref{S2}), 
\[\int_{\Gamma_{\emptyset}} \omega^{\color[rgb]{1.0,0.1,0.1}(1)}=S^{\color[rgb]{1,0.1,0}  (1)}(p^2,m^2) \quad \text{and}\quad
\int_{\Gamma_{\emptyset}} \omega^{\color[rgb]{.1,0.1,1.0}(2)}=S^{\color[rgb]{0.1,0.1,1}  (2)}(p^2,m^2),\] 
while the right entries are defined, respectively, by the two contours in eq.~(\ref{eq:homBasisSunset}). We used colour coding to emphasise the association of contour $\gamma_{1,2,3}^{{\color[rgb]{1.0,0.1,0.1}(1)}}$ to 
$\omega^{\color[rgb]{1.0,0.1,0.1}(1)}$ (red) and the contour $\gamma_{1,2,3}^{{\color[rgb]{0.1,0.1,1.0}(1)}}$ to 
$\omega^{\color[rgb]{0.1,0.1,1.0}(1)}$ (blue). This colour coding comes in handy in representing the coaction
 of eq.~(\ref{OneMassSunset_coaction}) diagrammatically:
\begin{equation}\label{eq:coactionS1}
\Delta\left[\begin{tikzpicture}[baseline={([yshift=-.5ex]current bounding box.center)}]
\coordinate (G1) at (0,0);
\coordinate (G2) at (1,0);
\coordinate (H1) at (-1/3,0);
\coordinate (H2) at (4/3,0);
\coordinate (I1) at (1/2,1/2);
\coordinate (I2) at (1/2,1/8);
\coordinate (I3) at (1/2,-1/2);
\coordinate (I4) at (1/2,-1/8);
\coordinate (J1) at (1,1/4);
\coordinate (J2) at (1,-1/4);
\coordinate (K1) at (1/2,-1/8);
\draw (G1) [line width=0.75 mm] -- (G2);
\draw (G1) to[out=80,in=100] (G2);
\draw (G1) to[out=-80,in=-100] (G2);
\draw (G1) [line width=0.75 mm] -- (H1);
\draw (G2) [line width=0.75 mm]-- (H2);
\node at (J1) [above=0 mm of J1] {\color{red}\small$(1)$};
\node at (J2) [below=0 mm of J2] {\vphantom{\small$(1)$}};
\end{tikzpicture}\right]=\begin{tikzpicture}[baseline={([yshift=-.5ex]current bounding box.center)}]
\coordinate (G1) at (0,0);
\coordinate (G2) at (1,0);
\coordinate (H1) at (-1/3,0);
\coordinate (H2) at (4/3,0);
\coordinate (I1) at (1/2,1/2);
\coordinate (I2) at (1/2,1/8);
\coordinate (I3) at (1/2,-1/2);
\coordinate (I4) at (1/2,-1/8);
\coordinate (J1) at (1,1/4);
\coordinate (J2) at (1,-1/4);
\coordinate (K1) at (1/2,-1/8);
\draw (G1) [line width=0.75 mm] -- (G2);
\draw (G1) to[out=80,in=100] (G2);
\draw (G1) to[out=-80,in=-100] (G2);
\draw (G1) [line width=0.75 mm] -- (H1);
\draw (G2) [line width=0.75 mm]-- (H2);
\node at (J1) [above=0 mm of J1] {\color{red}\small$(1)$};
\node at (J2) [below=0 mm of J2] {\vphantom{\small$(1)$}};
\end{tikzpicture}\otimes\begin{tikzpicture}[baseline={([yshift=-.5ex]current bounding box.center)}]
\coordinate (G1) at (0,0);
\coordinate (G2) at (1,0);
\coordinate (H1) at (-1/3,0);
\coordinate (H2) at (4/3,0);
\coordinate (I1) at (1/2,1/2);
\coordinate (I2) at (1/2,1/8);
\coordinate (I3) at (1/2,-1/2);
\coordinate (I4) at (1/2,-1/8);
\coordinate (J1) at (1,1/4);
\coordinate (J2) at (1,-1/4);
\coordinate (K1) at (1/2,-1/8);
\draw (G1) [line width=0.75 mm] -- (G2);
\draw (G1) to[out=80,in=100] (G2);
\draw (G1) to[out=-80,in=-100] (G2);
\draw (G1) [line width=0.75 mm] -- (H1);
\draw (G2) [line width=0.75 mm]-- (H2);
\draw (I1) [dashed,color=red,line width=0.5 mm] --(I3);
\node at (J1) [above=0 mm of J1] {\color{red}\small$(1)$};
\node at (J2) [below=0 mm of J2] {\vphantom{\small$(1)$}};
\end{tikzpicture}+\begin{tikzpicture}[baseline={([yshift=-.5ex]current bounding box.center)}]
\coordinate (G1) at (0,0);
\coordinate (G2) at (1,0);
\coordinate (H1) at (-1/3,0);
\coordinate (H2) at (4/3,0);
\coordinate (I1) at (1/2,1/2);
\coordinate (I2) at (1/2,1/8);
\coordinate (I3) at (1/2,-1/2);
\coordinate (I4) at (1/2,-1/8);
\coordinate (J1) at (1,1/4);
\coordinate (J2) at (1,-1/4);
\coordinate (K1) at (1/2,-1/8);
\draw (G1) [line width=0.75 mm] -- (G2);
\draw (G1) to[out=80,in=100] (G2);
\draw (G1) to[out=-80,in=-100] (G2);
\draw (G1) [line width=0.75 mm] -- (H1);
\draw (G2) [line width=0.75 mm]-- (H2);
\node at (J1) [above=0 mm of J1] {\color{blue}\small$(2)$};
\node at (J2) [below=0 mm of J2] {\vphantom{\small$(1)$}};
\end{tikzpicture}\otimes\begin{tikzpicture}[baseline={([yshift=-.5ex]current bounding box.center)}]
\coordinate (G1) at (0,0);
\coordinate (G2) at (1,0);
\coordinate (H1) at (-1/3,0);
\coordinate (H2) at (4/3,0);
\coordinate (I1) at (1/2,1/2);
\coordinate (I2) at (1/2,1/8);
\coordinate (I3) at (1/2,-1/2);
\coordinate (I4) at (1/2,-1/8);
\coordinate (J1) at (1,1/4);
\coordinate (J2) at (1,-1/4);
\coordinate (K1) at (1/2,-1/8);
\draw (G1) [line width=0.75 mm] -- (G2);
\draw (G1) to[out=80,in=100] (G2);
\draw (G1) to[out=-80,in=-100] (G2);
\draw (G1) [line width=0.75 mm] -- (H1);
\draw (G2) [line width=0.75 mm]-- (H2);
\draw (I1) [dashed,color=blue,line width=0.5 mm] --(I3);
\node at (J1) [above=0 mm of J1] {\color{red}\small$(1)$};
\node at (J2) [below=0 mm of J2] {\vphantom{\small$(1)$}};
\end{tikzpicture}\,,
\end{equation}
and
\begin{equation}\label{eq:coactionS2}
\Delta\left[\begin{tikzpicture}[baseline={([yshift=-.5ex]current bounding box.center)}]
\coordinate (G1) at (0,0);
\coordinate (G2) at (1,0);
\coordinate (H1) at (-1/3,0);
\coordinate (H2) at (4/3,0);
\coordinate (I1) at (1/2,1/2);
\coordinate (I2) at (1/2,1/8);
\coordinate (I3) at (1/2,-1/2);
\coordinate (I4) at (1/2,-1/8);
\coordinate (J1) at (1,1/4);
\coordinate (J2) at (1,-1/4);
\coordinate (K1) at (1/2,-1/8);
\draw (G1) [line width=0.75 mm] -- (G2);
\draw (G1) to[out=80,in=100] (G2);
\draw (G1) to[out=-80,in=-100] (G2);
\draw (G1) [line width=0.75 mm] -- (H1);
\draw (G2) [line width=0.75 mm]-- (H2);
\node at (J1) [above=0 mm of J1] {\color{blue}\small$(2)$};
\node at (J2) [below=0 mm of J2] {\vphantom{\small$(1)$}};
\end{tikzpicture}\right]=\begin{tikzpicture}[baseline={([yshift=-.5ex]current bounding box.center)}]
\coordinate (G1) at (0,0);
\coordinate (G2) at (1,0);
\coordinate (H1) at (-1/3,0);
\coordinate (H2) at (4/3,0);
\coordinate (I1) at (1/2,1/2);
\coordinate (I2) at (1/2,1/8);
\coordinate (I3) at (1/2,-1/2);
\coordinate (I4) at (1/2,-1/8);
\coordinate (J1) at (1,1/4);
\coordinate (J2) at (1,-1/4);
\coordinate (K1) at (1/2,-1/8);
\draw (G1) [line width=0.75 mm] -- (G2);
\draw (G1) to[out=80,in=100] (G2);
\draw (G1) to[out=-80,in=-100] (G2);
\draw (G1) [line width=0.75 mm] -- (H1);
\draw (G2) [line width=0.75 mm]-- (H2);
\node at (J1) [above=0 mm of J1] {\color{red}\small$(1)$};
\node at (J2) [below=0 mm of J2] {\vphantom{\small$(1)$}};
\end{tikzpicture}\otimes\begin{tikzpicture}[baseline={([yshift=-.5ex]current bounding box.center)}]
\coordinate (G1) at (0,0);
\coordinate (G2) at (1,0);
\coordinate (H1) at (-1/3,0);
\coordinate (H2) at (4/3,0);
\coordinate (I1) at (1/2,1/2);
\coordinate (I2) at (1/2,1/8);
\coordinate (I3) at (1/2,-1/2);
\coordinate (I4) at (1/2,-1/8);
\coordinate (J1) at (1,1/4);
\coordinate (J2) at (1,-1/4);
\coordinate (K1) at (1/2,-1/8);
\draw (G1) [line width=0.75 mm] -- (G2);
\draw (G1) to[out=80,in=100] (G2);
\draw (G1) to[out=-80,in=-100] (G2);
\draw (G1) [line width=0.75 mm] -- (H1);
\draw (G2) [line width=0.75 mm]-- (H2);
\draw (I1) [dashed,color=red,line width=0.5 mm] --(I3);
\node at (J1) [above=0 mm of J1] {\color{blue}\small$(2)$};
\node at (J2) [below=0 mm of J2] {\vphantom{\small$(1)$}};
\end{tikzpicture}+\begin{tikzpicture}[baseline={([yshift=-.5ex]current bounding box.center)}]
\coordinate (G1) at (0,0);
\coordinate (G2) at (1,0);
\coordinate (H1) at (-1/3,0);
\coordinate (H2) at (4/3,0);
\coordinate (I1) at (1/2,1/2);
\coordinate (I2) at (1/2,1/8);
\coordinate (I3) at (1/2,-1/2);
\coordinate (I4) at (1/2,-1/8);
\coordinate (J1) at (1,1/4);
\coordinate (J2) at (1,-1/4);
\coordinate (K1) at (1/2,-1/8);
\draw (G1) [line width=0.75 mm] -- (G2);
\draw (G1) to[out=80,in=100] (G2);
\draw (G1) to[out=-80,in=-100] (G2);
\draw (G1) [line width=0.75 mm] -- (H1);
\draw (G2) [line width=0.75 mm]-- (H2);
\node at (J1) [above=0 mm of J1] {\color{blue}\small$(2)$};
\node at (J2) [below=0 mm of J2] {\vphantom{\small$(1)$}};
\end{tikzpicture}\otimes\begin{tikzpicture}[baseline={([yshift=-.5ex]current bounding box.center)}]
\coordinate (G1) at (0,0);
\coordinate (G2) at (1,0);
\coordinate (H1) at (-1/3,0);
\coordinate (H2) at (4/3,0);
\coordinate (I1) at (1/2,1/2);
\coordinate (I2) at (1/2,1/8);
\coordinate (I3) at (1/2,-1/2);
\coordinate (I4) at (1/2,-1/8);
\coordinate (J1) at (1,1/4);
\coordinate (J2) at (1,-1/4);
\coordinate (K1) at (1/2,-1/8);
\draw (G1) [line width=0.75 mm] -- (G2);
\draw (G1) to[out=80,in=100] (G2);
\draw (G1) to[out=-80,in=-100] (G2);
\draw (G1) [line width=0.75 mm] -- (H1);
\draw (G2) [line width=0.75 mm]-- (H2);
\draw (I1) [dashed,color=blue,line width=0.5 mm] --(I3);
\node at (J1) [above=0 mm of J1] {\color{blue}\small$(2)$};
\node at (J2) [below=0 mm of J2] {\vphantom{\small$(1)$}};
\end{tikzpicture}\,.
\end{equation}
We have thus constructed the diagrammatic coaction for the one-mass sunset topology. Knowing the global coaction on the corresponding class of hypergeometric function made this task straightforward. The result, however, is non-trivial: it demonstrates that the diagrammatic coaction construction extends beyond one loop, despite the fact that the basis of master integrals is more complex, and consists of more than one integral with a given set of propagators. We emphasise that the result may be interpreted as a coaction on functions either as a global coaction, or as a local one, in which case it reproduces the coaction on MPLs to any order.
Of course, the example we have chosen here is simple. It has just two master integrals, both with all three propagators present, namely it does not contain any subtopologies (pinches of the original graph). This also implies that we need not consider any cuts except for the maximal cuts. Indeed, it can be shown that all non-vanishing cuts (as well as the uncut integral, barring $i\pi$ terms) can be expressed in this case in terms of the above maximal cut basis.
More complex topologies would have a rich structure of subtopologies, and corresponding cut contours where a subset of the propagators is put on-shell.  Ref.~\cite{Abreu:2021vhb}  explored that structure for a range of two-loop topologies. Here we illustrate some of the salient features using a couple of additional examples.

Our next example is the more general sunset integral with two non-vanishing
(and non-equal) internal masses. This topology is defined by 
\begin{align}\begin{split}\label{2mSunsetIntFamily}
&S(\nu_1,\nu_2,\nu_3,\nu_4,\nu_5;D;p^2,m_1^2,m_2^2)=\\
&\hspace*{40pt}=\left(\frac{e^{\gamma_E\epsilon}}{i\pi^{D/2}}\right)^2
\int d^Dk\int d^Dl\,
\frac{[m_2^2-(k+p)^2]^{-\nu_4}[m_1^2-(l+p)^2]^{-\nu_5}}
{[k^2-m_1^2]^{\nu_1}[l^2-m_2^2]^{\nu_2}[(k+l+p)^2]^{\nu_3}}\,,
\end{split}\end{align}
for integer $\nu_i$ with $\nu_4,\nu_5\leq0$ and for $D=n-2\epsilon$, with $n$ even.
There are four master integrals
in this topology, three of which are at the top topology sector, i.e.~featuring all three propagators, while 
the remaining one is the double tadpole with two massive propagators. The result for all these integrals and their cuts can be expressed in terms of Appell $F_4$ functions, which is significantly more involved than the Gauss hypergeometric function of the one-mass sunset case discussed above. The diagrammatic coaction can nevertheless be obtained following the steps described above, and using the coaction of the Appell $F_4$ functions derived in Ref.~\cite{Abreu:2019wzk}. The resulting diagrammatic coaction takes the form:
\begin{align}\begin{split}
\label{eq:Sunset2Mass1}
&\Delta\left[\begin{tikzpicture}[baseline={([yshift=-.5ex]current bounding box.center)}]
\coordinate (G1) at (0,0);
\coordinate (G2) at (1,0);
\coordinate (H1) at (-1/3,0);
\coordinate (H2) at (4/3,0);
\coordinate (I1) at (1/2,1/2);
\coordinate (I2) at (1/2,1/8);
\coordinate (I3) at (1/2,-1/2);
\coordinate (I4) at (1/2,-1/8);
\coordinate (J1) at (1,1/4);
\coordinate (J2) at (1,-1/4);
\coordinate (K1) at (1/2,-1/8);
\draw (G1) -- (G2);
\draw (G1) [line width=0.75 mm]to[out=80,in=100] (G2);
\draw (G1) [line width=0.75 mm]to[out=-80,in=-100] (G2);
\draw (G1) [line width=0.75 mm] -- (H1);
\draw (G2) [line width=0.75 mm]-- (H2);
\node at (J1) [above=0 mm of J1] {\color{orange}\small$(1)$};
\node at (J2) [below=0 mm of J2] {\vphantom{\small$(1)$}};
\end{tikzpicture}\right]
=\begin{tikzpicture}[baseline={([yshift=-.5ex]current bounding box.center)}]
\coordinate (G1) at (0,0);
\coordinate (G2) at (0,4/5);
\coordinate (G3) at (0,-4/5);
\coordinate (H1) at (-1/2,0);
\coordinate (H2) at (1/2,0);
\draw (G1) [line width=0.75 mm]to[out=120,in=180] (G2);
\draw (G2) [line width=0.75 mm]to[out=0,in=60] (G1);
\draw (G1) [line width=0.75 mm]to[out=-120,in=180] (G3);
\draw (G3) [line width=0.75 mm]to[out=0,in=-60] (G1);
\draw (G1) [line width=0.75 mm] -- (H1);
\draw (G1) [line width=0.75 mm]-- (H2);
\end{tikzpicture}\otimes
\left(
\begin{tikzpicture}[baseline={([yshift=-.5ex]current bounding box.center)}]
\coordinate (G1) at (0,0);
\coordinate (G2) at (1,0);
\coordinate (H1) at (-1/3,0);
\coordinate (H2) at (4/3,0);
\coordinate (I1) at (1/2,1/2);
\coordinate (I2) at (1/2,1/8);
\coordinate (I3) at (1/2,-1/2);
\coordinate (I4) at (1/2,-1/8);
\coordinate (J1) at (1,1/4);
\coordinate (J2) at (1,-1/4);
\coordinate (K1) at (1/2,-1/8);
\draw (G1) -- (G2);
\draw (G1) [line width=0.75 mm]to[out=80,in=100] (G2);
\draw (G1) [line width=0.75 mm]to[out=-80,in=-100] (G2);
\draw (G1) [line width=0.75 mm] -- (H1);
\draw (G2) [line width=0.75 mm]-- (H2);
\draw (I1) [dashed,color=red,line width=0.5 mm] --(I2);
\draw (I3) [dashed,color=red,line width=0.5 mm] --(I4);
\node at (J1) [above=0 mm of J1] {\color{orange}\small$(1)$};
\node at (J2) [below=0 mm of J2] {\vphantom{\small$(1)$}};
\end{tikzpicture}
+\begin{tikzpicture}[baseline={([yshift=-.5ex]current bounding box.center)}]
\coordinate (G1) at (0,0);
\coordinate (G2) at (1,0);
\coordinate (H1) at (-1/3,0);
\coordinate (H2) at (4/3,0);
\coordinate (I1) at (1/2,1/2);
\coordinate (I2) at (1/2,1/8);
\coordinate (I3) at (1/2,-1/2);
\coordinate (I4) at (1/2,-1/8);
\coordinate (J1) at (1,1/4);
\coordinate (J2) at (1,-1/4);
\coordinate (K1) at (1/2,-1/8);
\draw (G1) -- (G2);
\draw (G1) [line width=0.75 mm]to[out=80,in=100] (G2);
\draw (G1) [line width=0.75 mm]to[out=-80,in=-100] (G2);
\draw (G1) [line width=0.75 mm] -- (H1);
\draw (G2) [line width=0.75 mm]-- (H2);
\draw (I1) [dashed,color=orange,line width=0.5 mm] --(I3);
\node at (J1) [above=0 mm of J1] {\color{orange}\small$(1)$};
\node at (J2) [below=0 mm of J2] {\vphantom{\small$(1)$}};
\end{tikzpicture}+
\begin{tikzpicture}[baseline={([yshift=-.5ex]current bounding box.center)}]
\coordinate (G1) at (0,0);
\coordinate (G2) at (1,0);
\coordinate (H1) at (-1/3,0);
\coordinate (H2) at (4/3,0);
\coordinate (I1) at (1/2,1/2);
\coordinate (I2) at (1/2,1/8);
\coordinate (I3) at (1/2,-1/2);
\coordinate (I4) at (1/2,-1/8);
\coordinate (J1) at (1,1/4);
\coordinate (J2) at (1,-1/4);
\coordinate (K1) at (1/2,-1/8);
\draw (G1) -- (G2);
\draw (G1) [line width=0.75 mm]to[out=80,in=100] (G2);
\draw (G1) [line width=0.75 mm]to[out=-80,in=-100] (G2);
\draw (G1) [line width=0.75 mm] -- (H1);
\draw (G2) [line width=0.75 mm]-- (H2);
\draw (I1) [dashed,color=blue,line width=0.5 mm] --(I3);
\node at (J1) [above=0 mm of J1] {\color{orange}\small$(1)$};
\node at (J2) [below=0 mm of J2] {\vphantom{\small$(1)$}};
\end{tikzpicture}
+
\begin{tikzpicture}[baseline={([yshift=-.5ex]current bounding box.center)}]
\coordinate (G1) at (0,0);
\coordinate (G2) at (1,0);
\coordinate (H1) at (-1/3,0);
\coordinate (H2) at (4/3,0);
\coordinate (I1) at (1/2,1/2);
\coordinate (I2) at (1/2,1/8);
\coordinate (I3) at (1/2,-1/2);
\coordinate (I4) at (1/2,-1/8);
\coordinate (J1) at (1,1/4);
\coordinate (J2) at (1,-1/4);
\coordinate (K1) at (1/2,-1/8);
\draw (G1) -- (G2);
\draw (G1) [line width=0.75 mm]to[out=80,in=100] (G2);
\draw (G1) [line width=0.75 mm]to[out=-80,in=-100] (G2);
\draw (G1) [line width=0.75 mm] -- (H1);
\draw (G2) [line width=0.75 mm]-- (H2);
\draw (I1) [dashed,color=black!60!green,line width=0.5 mm] --(I3);
\node at (J1) [above=0 mm of J1] {\color{orange}\small$(1)$};
\node at (J2) [below=0 mm of J2] {\vphantom{\small$(1)$}};
\end{tikzpicture}
\right)\\
&+\,\begin{tikzpicture}[baseline={([yshift=-.5ex]current bounding box.center)}]
\coordinate (G1) at (0,0);
\coordinate (G2) at (1,0);
\coordinate (H1) at (-1/3,0);
\coordinate (H2) at (4/3,0);
\coordinate (I1) at (1/2,1/2);
\coordinate (I2) at (1/2,1/8);
\coordinate (I3) at (1/2,-1/2);
\coordinate (I4) at (1/2,-1/8);
\coordinate (J1) at (1,1/4);
\coordinate (J2) at (1,-1/4);
\coordinate (K1) at (1/2,-1/8);
\draw (G1) -- (G2);
\draw (G1) [line width=0.75 mm]to[out=80,in=100] (G2);
\draw (G1) [line width=0.75 mm]to[out=-80,in=-100] (G2);
\draw (G1) [line width=0.75 mm] -- (H1);
\draw (G2) [line width=0.75 mm]-- (H2);
\node at (J1) [above=0 mm of J1] {\color{orange}\small$(1)$};
\node at (J2) [below=0 mm of J2] {\vphantom{\small$(1)$}};
\end{tikzpicture}\otimes
\begin{tikzpicture}[baseline={([yshift=-.5ex]current bounding box.center)}]
\coordinate (G1) at (0,0);
\coordinate (G2) at (1,0);
\coordinate (H1) at (-1/3,0);
\coordinate (H2) at (4/3,0);
\coordinate (I1) at (1/2,1/2);
\coordinate (I2) at (1/2,1/8);
\coordinate (I3) at (1/2,-1/2);
\coordinate (I4) at (1/2,-1/8);
\coordinate (J1) at (1,1/4);
\coordinate (J2) at (1,-1/4);
\coordinate (K1) at (1/2,-1/8);
\draw (G1) -- (G2);
\draw (G1) [line width=0.75 mm]to[out=80,in=100] (G2);
\draw (G1) [line width=0.75 mm]to[out=-80,in=-100] (G2);
\draw (G1) [line width=0.75 mm] -- (H1);
\draw (G2) [line width=0.75 mm]-- (H2);
\draw (I1) [dashed,color=orange,line width=0.5 mm] --(I3);
\node at (J1) [above=0 mm of J1] {\color{orange}\small$(1)$};
\node at (J2) [below=0 mm of J2] {\vphantom{\small$(1)$}};
\end{tikzpicture}
+\begin{tikzpicture}[baseline={([yshift=-.5ex]current bounding box.center)}]
\coordinate (G1) at (0,0);
\coordinate (G2) at (1,0);
\coordinate (H1) at (-1/3,0);
\coordinate (H2) at (4/3,0);
\coordinate (I1) at (1/2,1/2);
\coordinate (I2) at (1/2,1/8);
\coordinate (I3) at (1/2,-1/2);
\coordinate (I4) at (1/2,-1/8);
\coordinate (J1) at (1,1/4);
\coordinate (J2) at (1,-1/4);
\coordinate (K1) at (1/2,-1/8);
\draw (G1) -- (G2);
\draw (G1) [line width=0.75 mm]to[out=80,in=100] (G2);
\draw (G1) [line width=0.75 mm]to[out=-80,in=-100] (G2);
\draw (G1) [line width=0.75 mm] -- (H1);
\draw (G2) [line width=0.75 mm]-- (H2);
\node at (J1) [above=0 mm of J1] {\color{blue}\small$(2)$};
\node at (J2) [below=0 mm of J2] {\vphantom{\small$(2)$}};
\end{tikzpicture}\otimes
\begin{tikzpicture}[baseline={([yshift=-.5ex]current bounding box.center)}]
\coordinate (G1) at (0,0);
\coordinate (G2) at (1,0);
\coordinate (H1) at (-1/3,0);
\coordinate (H2) at (4/3,0);
\coordinate (I1) at (1/2,1/2);
\coordinate (I2) at (1/2,1/8);
\coordinate (I3) at (1/2,-1/2);
\coordinate (I4) at (1/2,-1/8);
\coordinate (J1) at (1,1/4);
\coordinate (J2) at (1,-1/4);
\coordinate (K1) at (1/2,-1/8);
\draw (G1) -- (G2);
\draw (G1) [line width=0.75 mm]to[out=80,in=100] (G2);
\draw (G1) [line width=0.75 mm]to[out=-80,in=-100] (G2);
\draw (G1) [line width=0.75 mm] -- (H1);
\draw (G2) [line width=0.75 mm]-- (H2);
\draw (I1) [dashed,color=blue,line width=0.5 mm] --(I3);
\node at (J1) [above=0 mm of J1] {\color{orange}\small$(1)$};
\node at (J2) [below=0 mm of J2] {\vphantom{\small$(1)$}};
\end{tikzpicture}
+\begin{tikzpicture}[baseline={([yshift=-.5ex]current bounding box.center)}]
\coordinate (G1) at (0,0);
\coordinate (G2) at (1,0);
\coordinate (H1) at (-1/3,0);
\coordinate (H2) at (4/3,0);
\coordinate (I1) at (1/2,1/2);
\coordinate (I2) at (1/2,1/8);
\coordinate (I3) at (1/2,-1/2);
\coordinate (I4) at (1/2,-1/8);
\coordinate (J1) at (1,1/4);
\coordinate (J2) at (1,-1/4);
\coordinate (K1) at (1/2,-1/8);
\draw (G1) -- (G2);
\draw (G1) [line width=0.75 mm]to[out=80,in=100] (G2);
\draw (G1) [line width=0.75 mm]to[out=-80,in=-100] (G2);
\draw (G1) [line width=0.75 mm] -- (H1);
\draw (G2) [line width=0.75 mm]-- (H2);
\node at (J1) [above=0 mm of J1] {\color{black!60!green}\small$(3)$};
\node at (J2) [below=0 mm of J2] {\vphantom{\small$(3)$}};
\end{tikzpicture}\otimes
\begin{tikzpicture}[baseline={([yshift=-.5ex]current bounding box.center)}]
\coordinate (G1) at (0,0);
\coordinate (G2) at (1,0);
\coordinate (H1) at (-1/3,0);
\coordinate (H2) at (4/3,0);
\coordinate (I1) at (1/2,1/2);
\coordinate (I2) at (1/2,1/8);
\coordinate (I3) at (1/2,-1/2);
\coordinate (I4) at (1/2,-1/8);
\coordinate (J1) at (1,1/4);
\coordinate (J2) at (1,-1/4);
\coordinate (K1) at (1/2,-1/8);
\draw (G1) -- (G2);
\draw (G1) [line width=0.75 mm]to[out=80,in=100] (G2);
\draw (G1) [line width=0.75 mm]to[out=-80,in=-100] (G2);
\draw (G1) [line width=0.75 mm] -- (H1);
\draw (G2) [line width=0.75 mm]-- (H2);
\draw (I1) [dashed,color=black!60!green,line width=0.5 mm] --(I3);
\node at (J1) [above=0 mm of J1] {\color{orange}\small$(1)$};
\node at (J2) [below=0 mm of J2] {\vphantom{\small$(1)$}};
\end{tikzpicture}\,,
\end{split}
\end{align}
%%%%%%%%%%%%%%%%%%%%%%%%%%%%%%%%%
%
%
%		NEW INTEGRAL!!!
%
%
%%%%%%%%%%%%%%%%%%%%%%%%%%%%%%%%%
\begin{align}\begin{split}
\label{eq:Sunset2Mass2}
&\Delta\left[\begin{tikzpicture}[baseline={([yshift=-.5ex]current bounding box.center)}]
\coordinate (G1) at (0,0);
\coordinate (G2) at (1,0);
\coordinate (H1) at (-1/3,0);
\coordinate (H2) at (4/3,0);
\coordinate (I1) at (1/2,1/2);
\coordinate (I2) at (1/2,1/8);
\coordinate (I3) at (1/2,-1/2);
\coordinate (I4) at (1/2,-1/8);
\coordinate (J1) at (1,1/4);
\coordinate (J2) at (1,-1/4);
\coordinate (K1) at (1/2,-1/8);
\draw (G1) -- (G2);
\draw (G1) [line width=0.75 mm]to[out=80,in=100] (G2);
\draw (G1) [line width=0.75 mm]to[out=-80,in=-100] (G2);
\draw (G1) [line width=0.75 mm] -- (H1);
\draw (G2) [line width=0.75 mm]-- (H2);
\node at (J1) [above=0 mm of J1] {\color{blue}\small$(2)$};
\node at (J2) [below=0 mm of J2] {\vphantom{\small$(2)$}};
\end{tikzpicture}\right]
=\begin{tikzpicture}[baseline={([yshift=-.5ex]current bounding box.center)}]
\coordinate (G1) at (0,0);
\coordinate (G2) at (0,4/5);
\coordinate (G3) at (0,-4/5);
\coordinate (H1) at (-1/2,0);
\coordinate (H2) at (1/2,0);
\draw (G1) [line width=0.75 mm]to[out=120,in=180] (G2);
\draw (G2) [line width=0.75 mm]to[out=0,in=60] (G1);
\draw (G1) [line width=0.75 mm]to[out=-120,in=180] (G3);
\draw (G3) [line width=0.75 mm]to[out=0,in=-60] (G1);
\draw (G1) [line width=0.75 mm] -- (H1);
\draw (G1) [line width=0.75 mm]-- (H2);
\end{tikzpicture}\otimes
\left(
\begin{tikzpicture}[baseline={([yshift=-.5ex]current bounding box.center)}]
\coordinate (G1) at (0,0);
\coordinate (G2) at (1,0);
\coordinate (H1) at (-1/3,0);
\coordinate (H2) at (4/3,0);
\coordinate (I1) at (1/2,1/2);
\coordinate (I2) at (1/2,1/8);
\coordinate (I3) at (1/2,-1/2);
\coordinate (I4) at (1/2,-1/8);
\coordinate (J1) at (1,1/4);
\coordinate (J2) at (1,-1/4);
\coordinate (K1) at (1/2,-1/8);
\draw (G1) -- (G2);
\draw (G1) [line width=0.75 mm]to[out=80,in=100] (G2);
\draw (G1) [line width=0.75 mm]to[out=-80,in=-100] (G2);
\draw (G1) [line width=0.75 mm] -- (H1);
\draw (G2) [line width=0.75 mm]-- (H2);
\draw (I1) [dashed,color=red,line width=0.5 mm] --(I2);
\draw (I3) [dashed,color=red,line width=0.5 mm] --(I4);
\node at (J1) [above=0 mm of J1] {\color{blue}\small$(2)$};
\node at (J2) [below=0 mm of J2] {\vphantom{\small$(2)$}};
\end{tikzpicture}
+\begin{tikzpicture}[baseline={([yshift=-.5ex]current bounding box.center)}]
\coordinate (G1) at (0,0);
\coordinate (G2) at (1,0);
\coordinate (H1) at (-1/3,0);
\coordinate (H2) at (4/3,0);
\coordinate (I1) at (1/2,1/2);
\coordinate (I2) at (1/2,1/8);
\coordinate (I3) at (1/2,-1/2);
\coordinate (I4) at (1/2,-1/8);
\coordinate (J1) at (1,1/4);
\coordinate (J2) at (1,-1/4);
\coordinate (K1) at (1/2,-1/8);
\draw (G1) -- (G2);
\draw (G1) [line width=0.75 mm]to[out=80,in=100] (G2);
\draw (G1) [line width=0.75 mm]to[out=-80,in=-100] (G2);
\draw (G1) [line width=0.75 mm] -- (H1);
\draw (G2) [line width=0.75 mm]-- (H2);
\draw (I1) [dashed,color=orange,line width=0.5 mm] --(I3);
\node at (J1) [above=0 mm of J1] {\color{blue}\small$(2)$};
\node at (J2) [below=0 mm of J2] {\vphantom{\small$(2)$}};
\end{tikzpicture}+
\begin{tikzpicture}[baseline={([yshift=-.5ex]current bounding box.center)}]
\coordinate (G1) at (0,0);
\coordinate (G2) at (1,0);
\coordinate (H1) at (-1/3,0);
\coordinate (H2) at (4/3,0);
\coordinate (I1) at (1/2,1/2);
\coordinate (I2) at (1/2,1/8);
\coordinate (I3) at (1/2,-1/2);
\coordinate (I4) at (1/2,-1/8);
\coordinate (J1) at (1,1/4);
\coordinate (J2) at (1,-1/4);
\coordinate (K1) at (1/2,-1/8);
\draw (G1) -- (G2);
\draw (G1) [line width=0.75 mm]to[out=80,in=100] (G2);
\draw (G1) [line width=0.75 mm]to[out=-80,in=-100] (G2);
\draw (G1) [line width=0.75 mm] -- (H1);
\draw (G2) [line width=0.75 mm]-- (H2);
\draw (I1) [dashed,color=blue,line width=0.5 mm] --(I3);
\node at (J1) [above=0 mm of J1] {\color{blue}\small$(2)$};
\node at (J2) [below=0 mm of J2] {\vphantom{\small$(2)$}};
\end{tikzpicture}
+
\begin{tikzpicture}[baseline={([yshift=-.5ex]current bounding box.center)}]
\coordinate (G1) at (0,0);
\coordinate (G2) at (1,0);
\coordinate (H1) at (-1/3,0);
\coordinate (H2) at (4/3,0);
\coordinate (I1) at (1/2,1/2);
\coordinate (I2) at (1/2,1/8);
\coordinate (I3) at (1/2,-1/2);
\coordinate (I4) at (1/2,-1/8);
\coordinate (J1) at (1,1/4);
\coordinate (J2) at (1,-1/4);
\coordinate (K1) at (1/2,-1/8);
\draw (G1) -- (G2);
\draw (G1) [line width=0.75 mm]to[out=80,in=100] (G2);
\draw (G1) [line width=0.75 mm]to[out=-80,in=-100] (G2);
\draw (G1) [line width=0.75 mm] -- (H1);
\draw (G2) [line width=0.75 mm]-- (H2);
\draw (I1) [dashed,color=black!60!green,line width=0.5 mm] --(I3);
\node at (J1) [above=0 mm of J1] {\color{blue}\small$(2)$};
\node at (J2) [below=0 mm of J2] {\vphantom{\small$(2)$}};
\end{tikzpicture}
\right)\\
&+\,\begin{tikzpicture}[baseline={([yshift=-.5ex]current bounding box.center)}]
\coordinate (G1) at (0,0);
\coordinate (G2) at (1,0);
\coordinate (H1) at (-1/3,0);
\coordinate (H2) at (4/3,0);
\coordinate (I1) at (1/2,1/2);
\coordinate (I2) at (1/2,1/8);
\coordinate (I3) at (1/2,-1/2);
\coordinate (I4) at (1/2,-1/8);
\coordinate (J1) at (1,1/4);
\coordinate (J2) at (1,-1/4);
\coordinate (K1) at (1/2,-1/8);
\draw (G1) -- (G2);
\draw (G1) [line width=0.75 mm]to[out=80,in=100] (G2);
\draw (G1) [line width=0.75 mm]to[out=-80,in=-100] (G2);
\draw (G1) [line width=0.75 mm] -- (H1);
\draw (G2) [line width=0.75 mm]-- (H2);
\node at (J1) [above=0 mm of J1] {\color{orange}\small$(1)$};
\node at (J2) [below=0 mm of J2] {\vphantom{\small$(1)$}};
\end{tikzpicture}\otimes
\begin{tikzpicture}[baseline={([yshift=-.5ex]current bounding box.center)}]
\coordinate (G1) at (0,0);
\coordinate (G2) at (1,0);
\coordinate (H1) at (-1/3,0);
\coordinate (H2) at (4/3,0);
\coordinate (I1) at (1/2,1/2);
\coordinate (I2) at (1/2,1/8);
\coordinate (I3) at (1/2,-1/2);
\coordinate (I4) at (1/2,-1/8);
\coordinate (J1) at (1,1/4);
\coordinate (J2) at (1,-1/4);
\coordinate (K1) at (1/2,-1/8);
\draw (G1) -- (G2);
\draw (G1) [line width=0.75 mm]to[out=80,in=100] (G2);
\draw (G1) [line width=0.75 mm]to[out=-80,in=-100] (G2);
\draw (G1) [line width=0.75 mm] -- (H1);
\draw (G2) [line width=0.75 mm]-- (H2);
\draw (I1) [dashed,color=orange,line width=0.5 mm] --(I3);
\node at (J1) [above=0 mm of J1] {\color{blue}\small$(2)$};
\node at (J2) [below=0 mm of J2] {\vphantom{\small$(2)$}};
\end{tikzpicture}
+\begin{tikzpicture}[baseline={([yshift=-.5ex]current bounding box.center)}]
\coordinate (G1) at (0,0);
\coordinate (G2) at (1,0);
\coordinate (H1) at (-1/3,0);
\coordinate (H2) at (4/3,0);
\coordinate (I1) at (1/2,1/2);
\coordinate (I2) at (1/2,1/8);
\coordinate (I3) at (1/2,-1/2);
\coordinate (I4) at (1/2,-1/8);
\coordinate (J1) at (1,1/4);
\coordinate (J2) at (1,-1/4);
\coordinate (K1) at (1/2,-1/8);
\draw (G1) -- (G2);
\draw (G1) [line width=0.75 mm]to[out=80,in=100] (G2);
\draw (G1) [line width=0.75 mm]to[out=-80,in=-100] (G2);
\draw (G1) [line width=0.75 mm] -- (H1);
\draw (G2) [line width=0.75 mm]-- (H2);
\node at (J1) [above=0 mm of J1] {\color{blue}\small$(2)$};
\node at (J2) [below=0 mm of J2] {\vphantom{\small$(2)$}};
\end{tikzpicture}\otimes
\begin{tikzpicture}[baseline={([yshift=-.5ex]current bounding box.center)}]
\coordinate (G1) at (0,0);
\coordinate (G2) at (1,0);
\coordinate (H1) at (-1/3,0);
\coordinate (H2) at (4/3,0);
\coordinate (I1) at (1/2,1/2);
\coordinate (I2) at (1/2,1/8);
\coordinate (I3) at (1/2,-1/2);
\coordinate (I4) at (1/2,-1/8);
\coordinate (J1) at (1,1/4);
\coordinate (J2) at (1,-1/4);
\coordinate (K1) at (1/2,-1/8);
\draw (G1) -- (G2);
\draw (G1) [line width=0.75 mm]to[out=80,in=100] (G2);
\draw (G1) [line width=0.75 mm]to[out=-80,in=-100] (G2);
\draw (G1) [line width=0.75 mm] -- (H1);
\draw (G2) [line width=0.75 mm]-- (H2);
\draw (I1) [dashed,color=blue,line width=0.5 mm] --(I3);
\node at (J1) [above=0 mm of J1] {\color{blue}\small$(2)$};
\node at (J2) [below=0 mm of J2] {\vphantom{\small$(2)$}};
\end{tikzpicture}
+\begin{tikzpicture}[baseline={([yshift=-.5ex]current bounding box.center)}]
\coordinate (G1) at (0,0);
\coordinate (G2) at (1,0);
\coordinate (H1) at (-1/3,0);
\coordinate (H2) at (4/3,0);
\coordinate (I1) at (1/2,1/2);
\coordinate (I2) at (1/2,1/8);
\coordinate (I3) at (1/2,-1/2);
\coordinate (I4) at (1/2,-1/8);
\coordinate (J1) at (1,1/4);
\coordinate (J2) at (1,-1/4);
\coordinate (K1) at (1/2,-1/8);
\draw (G1) -- (G2);
\draw (G1) [line width=0.75 mm]to[out=80,in=100] (G2);
\draw (G1) [line width=0.75 mm]to[out=-80,in=-100] (G2);
\draw (G1) [line width=0.75 mm] -- (H1);
\draw (G2) [line width=0.75 mm]-- (H2);
\node at (J1) [above=0 mm of J1] {\color{black!60!green}\small$(3)$};
\node at (J2) [below=0 mm of J2] {\vphantom{\small$(3)$}};
\end{tikzpicture}\otimes
\begin{tikzpicture}[baseline={([yshift=-.5ex]current bounding box.center)}]
\coordinate (G1) at (0,0);
\coordinate (G2) at (1,0);
\coordinate (H1) at (-1/3,0);
\coordinate (H2) at (4/3,0);
\coordinate (I1) at (1/2,1/2);
\coordinate (I2) at (1/2,1/8);
\coordinate (I3) at (1/2,-1/2);
\coordinate (I4) at (1/2,-1/8);
\coordinate (J1) at (1,1/4);
\coordinate (J2) at (1,-1/4);
\coordinate (K1) at (1/2,-1/8);
\draw (G1) -- (G2);
\draw (G1) [line width=0.75 mm]to[out=80,in=100] (G2);
\draw (G1) [line width=0.75 mm]to[out=-80,in=-100] (G2);
\draw (G1) [line width=0.75 mm] -- (H1);
\draw (G2) [line width=0.75 mm]-- (H2);
\draw (I1) [dashed,color=black!60!green,line width=0.5 mm] --(I3);
\node at (J1) [above=0 mm of J1] {\color{blue}\small$(2)$};
\node at (J2) [below=0 mm of J2] {\vphantom{\small$(2)$}};
\end{tikzpicture}\,,
\end{split}
\end{align}
%%%%%%%%%%%%%%%%%%%%%%%%%%%%%%%%%
%
%
%		NEW INTEGRAL!!!
%
%
%%%%%%%%%%%%%%%%%%%%%%%%%%%%%%%%%
\begin{align}
\begin{split}
\label{eq:Sunset2Mass3}
&\Delta\left[\begin{tikzpicture}[baseline={([yshift=-.5ex]current bounding box.center)}]
\coordinate (G1) at (0,0);
\coordinate (G2) at (1,0);
\coordinate (H1) at (-1/3,0);
\coordinate (H2) at (4/3,0);
\coordinate (I1) at (1/2,1/2);
\coordinate (I2) at (1/2,1/8);
\coordinate (I3) at (1/2,-1/2);
\coordinate (I4) at (1/2,-1/8);
\coordinate (J1) at (1,1/4);
\coordinate (J2) at (1,-1/4);
\coordinate (K1) at (1/2,-1/8);
\draw (G1) -- (G2);
\draw (G1) [line width=0.75 mm]to[out=80,in=100] (G2);
\draw (G1) [line width=0.75 mm]to[out=-80,in=-100] (G2);
\draw (G1) [line width=0.75 mm] -- (H1);
\draw (G2) [line width=0.75 mm]-- (H2);
\node at (J1) [above=0 mm of J1] {\color{black!60!green}\small$(3)$};
\node at (J2) [below=0 mm of J2] {\vphantom{\small$(3)$}};
\end{tikzpicture}\right]
=\begin{tikzpicture}[baseline={([yshift=-.5ex]current bounding box.center)}]
\coordinate (G1) at (0,0);
\coordinate (G2) at (0,4/5);
\coordinate (G3) at (0,-4/5);
\coordinate (H1) at (-1/2,0);
\coordinate (H2) at (1/2,0);
\draw (G1) [line width=0.75 mm]to[out=120,in=180] (G2);
\draw (G2) [line width=0.75 mm]to[out=0,in=60] (G1);
\draw (G1) [line width=0.75 mm]to[out=-120,in=180] (G3);
\draw (G3) [line width=0.75 mm]to[out=0,in=-60] (G1);
\draw (G1) [line width=0.75 mm] -- (H1);
\draw (G1) [line width=0.75 mm]-- (H2);
\end{tikzpicture}\otimes
\left(
\begin{tikzpicture}[baseline={([yshift=-.5ex]current bounding box.center)}]
\coordinate (G1) at (0,0);
\coordinate (G2) at (1,0);
\coordinate (H1) at (-1/3,0);
\coordinate (H2) at (4/3,0);
\coordinate (I1) at (1/2,1/2);
\coordinate (I2) at (1/2,1/8);
\coordinate (I3) at (1/2,-1/2);
\coordinate (I4) at (1/2,-1/8);
\coordinate (J1) at (1,1/4);
\coordinate (J2) at (1,-1/4);
\coordinate (K1) at (1/2,-1/8);
\draw (G1) -- (G2);
\draw (G1) [line width=0.75 mm]to[out=80,in=100] (G2);
\draw (G1) [line width=0.75 mm]to[out=-80,in=-100] (G2);
\draw (G1) [line width=0.75 mm] -- (H1);
\draw (G2) [line width=0.75 mm]-- (H2);
\draw (I1) [dashed,color=red,line width=0.5 mm] --(I2);
\draw (I3) [dashed,color=red,line width=0.5 mm] --(I4);
\node at (J1) [above=0 mm of J1] {\color{black!60!green}\small$(3)$};
\node at (J2) [below=0 mm of J2] {\vphantom{\small$(3)$}};
\end{tikzpicture}
+\begin{tikzpicture}[baseline={([yshift=-.5ex]current bounding box.center)}]
\coordinate (G1) at (0,0);
\coordinate (G2) at (1,0);
\coordinate (H1) at (-1/3,0);
\coordinate (H2) at (4/3,0);
\coordinate (I1) at (1/2,1/2);
\coordinate (I2) at (1/2,1/8);
\coordinate (I3) at (1/2,-1/2);
\coordinate (I4) at (1/2,-1/8);
\coordinate (J1) at (1,1/4);
\coordinate (J2) at (1,-1/4);
\coordinate (K1) at (1/2,-1/8);
\draw (G1) -- (G2);
\draw (G1) [line width=0.75 mm]to[out=80,in=100] (G2);
\draw (G1) [line width=0.75 mm]to[out=-80,in=-100] (G2);
\draw (G1) [line width=0.75 mm] -- (H1);
\draw (G2) [line width=0.75 mm]-- (H2);
\draw (I1) [dashed,color=orange,line width=0.5 mm] --(I3);
\node at (J1) [above=0 mm of J1] {\color{black!60!green}\small$(3)$};
\node at (J2) [below=0 mm of J2] {\vphantom{\small$(3)$}};
\end{tikzpicture}+
\begin{tikzpicture}[baseline={([yshift=-.5ex]current bounding box.center)}]
\coordinate (G1) at (0,0);
\coordinate (G2) at (1,0);
\coordinate (H1) at (-1/3,0);
\coordinate (H2) at (4/3,0);
\coordinate (I1) at (1/2,1/2);
\coordinate (I2) at (1/2,1/8);
\coordinate (I3) at (1/2,-1/2);
\coordinate (I4) at (1/2,-1/8);
\coordinate (J1) at (1,1/4);
\coordinate (J2) at (1,-1/4);
\coordinate (K1) at (1/2,-1/8);
\draw (G1) -- (G2);
\draw (G1) [line width=0.75 mm]to[out=80,in=100] (G2);
\draw (G1) [line width=0.75 mm]to[out=-80,in=-100] (G2);
\draw (G1) [line width=0.75 mm] -- (H1);
\draw (G2) [line width=0.75 mm]-- (H2);
\draw (I1) [dashed,color=blue,line width=0.5 mm] --(I3);
\node at (J1) [above=0 mm of J1] {\color{black!60!green}\small$(3)$};
\node at (J2) [below=0 mm of J2] {\vphantom{\small$(3)$}};
\end{tikzpicture}
+
\begin{tikzpicture}[baseline={([yshift=-.5ex]current bounding box.center)}]
\coordinate (G1) at (0,0);
\coordinate (G2) at (1,0);
\coordinate (H1) at (-1/3,0);
\coordinate (H2) at (4/3,0);
\coordinate (I1) at (1/2,1/2);
\coordinate (I2) at (1/2,1/8);
\coordinate (I3) at (1/2,-1/2);
\coordinate (I4) at (1/2,-1/8);
\coordinate (J1) at (1,1/4);
\coordinate (J2) at (1,-1/4);
\coordinate (K1) at (1/2,-1/8);
\draw (G1) -- (G2);
\draw (G1) [line width=0.75 mm]to[out=80,in=100] (G2);
\draw (G1) [line width=0.75 mm]to[out=-80,in=-100] (G2);
\draw (G1) [line width=0.75 mm] -- (H1);
\draw (G2) [line width=0.75 mm]-- (H2);
\draw (I1) [dashed,color=black!60!green,line width=0.5 mm] --(I3);
\node at (J1) [above=0 mm of J1] {\color{black!60!green}\small$(3)$};
\node at (J2) [below=0 mm of J2] {\vphantom{\small$(3)$}};
\end{tikzpicture}
\right)\\
&+\,\begin{tikzpicture}[baseline={([yshift=-.5ex]current bounding box.center)}]
\coordinate (G1) at (0,0);
\coordinate (G2) at (1,0);
\coordinate (H1) at (-1/3,0);
\coordinate (H2) at (4/3,0);
\coordinate (I1) at (1/2,1/2);
\coordinate (I2) at (1/2,1/8);
\coordinate (I3) at (1/2,-1/2);
\coordinate (I4) at (1/2,-1/8);
\coordinate (J1) at (1,1/4);
\coordinate (J2) at (1,-1/4);
\coordinate (K1) at (1/2,-1/8);
\draw (G1) -- (G2);
\draw (G1) [line width=0.75 mm]to[out=80,in=100] (G2);
\draw (G1) [line width=0.75 mm]to[out=-80,in=-100] (G2);
\draw (G1) [line width=0.75 mm] -- (H1);
\draw (G2) [line width=0.75 mm]-- (H2);
\node at (J1) [above=0 mm of J1] {\color{orange}\small$(1)$};
\node at (J2) [below=0 mm of J2] {\vphantom{\small$(1)$}};
\end{tikzpicture}\otimes
\begin{tikzpicture}[baseline={([yshift=-.5ex]current bounding box.center)}]
\coordinate (G1) at (0,0);
\coordinate (G2) at (1,0);
\coordinate (H1) at (-1/3,0);
\coordinate (H2) at (4/3,0);
\coordinate (I1) at (1/2,1/2);
\coordinate (I2) at (1/2,1/8);
\coordinate (I3) at (1/2,-1/2);
\coordinate (I4) at (1/2,-1/8);
\coordinate (J1) at (1,1/4);
\coordinate (J2) at (1,-1/4);
\coordinate (K1) at (1/2,-1/8);
\draw (G1) -- (G2);
\draw (G1) [line width=0.75 mm]to[out=80,in=100] (G2);
\draw (G1) [line width=0.75 mm]to[out=-80,in=-100] (G2);
\draw (G1) [line width=0.75 mm] -- (H1);
\draw (G2) [line width=0.75 mm]-- (H2);
\draw (I1) [dashed,color=orange,line width=0.5 mm] --(I3);
\node at (J1) [above=0 mm of J1] {\color{black!60!green}\small$(3)$};
\node at (J2) [below=0 mm of J2] {\vphantom{\small$(3)$}};
\end{tikzpicture}
+\begin{tikzpicture}[baseline={([yshift=-.5ex]current bounding box.center)}]
\coordinate (G1) at (0,0);
\coordinate (G2) at (1,0);
\coordinate (H1) at (-1/3,0);
\coordinate (H2) at (4/3,0);
\coordinate (I1) at (1/2,1/2);
\coordinate (I2) at (1/2,1/8);
\coordinate (I3) at (1/2,-1/2);
\coordinate (I4) at (1/2,-1/8);
\coordinate (J1) at (1,1/4);
\coordinate (J2) at (1,-1/4);
\coordinate (K1) at (1/2,-1/8);
\draw (G1) -- (G2);
\draw (G1) [line width=0.75 mm]to[out=80,in=100] (G2);
\draw (G1) [line width=0.75 mm]to[out=-80,in=-100] (G2);
\draw (G1) [line width=0.75 mm] -- (H1);
\draw (G2) [line width=0.75 mm]-- (H2);
\node at (J1) [above=0 mm of J1] {\color{blue}\small$(2)$};
\node at (J2) [below=0 mm of J2] {\vphantom{\small$(2)$}};
\end{tikzpicture}\otimes
\begin{tikzpicture}[baseline={([yshift=-.5ex]current bounding box.center)}]
\coordinate (G1) at (0,0);
\coordinate (G2) at (1,0);
\coordinate (H1) at (-1/3,0);
\coordinate (H2) at (4/3,0);
\coordinate (I1) at (1/2,1/2);
\coordinate (I2) at (1/2,1/8);
\coordinate (I3) at (1/2,-1/2);
\coordinate (I4) at (1/2,-1/8);
\coordinate (J1) at (1,1/4);
\coordinate (J2) at (1,-1/4);
\coordinate (K1) at (1/2,-1/8);
\draw (G1) -- (G2);
\draw (G1) [line width=0.75 mm]to[out=80,in=100] (G2);
\draw (G1) [line width=0.75 mm]to[out=-80,in=-100] (G2);
\draw (G1) [line width=0.75 mm] -- (H1);
\draw (G2) [line width=0.75 mm]-- (H2);
\draw (I1) [dashed,color=blue,line width=0.5 mm] --(I3);
\node at (J1) [above=0 mm of J1] {\color{black!60!green}\small$(3)$};
\node at (J2) [below=0 mm of J2] {\vphantom{\small$(3)$}};
\end{tikzpicture}
+\begin{tikzpicture}[baseline={([yshift=-.5ex]current bounding box.center)}]
\coordinate (G1) at (0,0);
\coordinate (G2) at (1,0);
\coordinate (H1) at (-1/3,0);
\coordinate (H2) at (4/3,0);
\coordinate (I1) at (1/2,1/2);
\coordinate (I2) at (1/2,1/8);
\coordinate (I3) at (1/2,-1/2);
\coordinate (I4) at (1/2,-1/8);
\coordinate (J1) at (1,1/4);
\coordinate (J2) at (1,-1/4);
\coordinate (K1) at (1/2,-1/8);
\draw (G1) -- (G2);
\draw (G1) [line width=0.75 mm]to[out=80,in=100] (G2);
\draw (G1) [line width=0.75 mm]to[out=-80,in=-100] (G2);
\draw (G1) [line width=0.75 mm] -- (H1);
\draw (G2) [line width=0.75 mm]-- (H2);
\node at (J1) [above=0 mm of J1] {\color{black!60!green}\small$(3)$};
\node at (J2) [below=0 mm of J2] {\vphantom{\small$(3)$}};
\end{tikzpicture}\otimes
\begin{tikzpicture}[baseline={([yshift=-.5ex]current bounding box.center)}]
\coordinate (G1) at (0,0);
\coordinate (G2) at (1,0);
\coordinate (H1) at (-1/3,0);
\coordinate (H2) at (4/3,0);
\coordinate (I1) at (1/2,1/2);
\coordinate (I2) at (1/2,1/8);
\coordinate (I3) at (1/2,-1/2);
\coordinate (I4) at (1/2,-1/8);
\coordinate (J1) at (1,1/4);
\coordinate (J2) at (1,-1/4);
\coordinate (K1) at (1/2,-1/8);
\draw (G1) -- (G2);
\draw (G1) [line width=0.75 mm]to[out=80,in=100] (G2);
\draw (G1) [line width=0.75 mm]to[out=-80,in=-100] (G2);
\draw (G1) [line width=0.75 mm] -- (H1);
\draw (G2) [line width=0.75 mm]-- (H2);
\draw (I1) [dashed,color=black!60!green,line width=0.5 mm] --(I3);
\node at (J1) [above=0 mm of J1] {\color{black!60!green}\small$(3)$};
\node at (J2) [below=0 mm of J2] {\vphantom{\small$(3)$}};
\end{tikzpicture}\,,
\end{split}
\end{align}
where we associate the colours orange, blue and green to the three master integrals and the corresponding cuts.
Note that similarly to the one-loop case, the contour dual to the double tadpole consists not only of the two-propagator cut but also of additional maximal-cut terms. 

As a final example we choose to present the three-mass double-edged triangle with massless propagators. This topology is defined by 
\begin{align}\begin{split}\label{eq:doubleEdgedGen}
P(\nu_1,\nu_2,\nu_3,\nu_4,&\,\nu_5,\nu_6,\nu_7;D;p_1^2,p_2^2,p_3^2)\\
=&\left(\frac{e^{\gamma_E\epsilon}}{i\pi^{D/2}}\right)^2
\int d^{D}l\int d^{D}k
\frac{[(k+p_3)^2]^{-\nu_5}[(k+p_2)^2]^{-\nu_6}[(l+p_2)^2]^{-\nu_7}}
{[k^2]^{\nu_1}[(k+l+p_2)^2]^{\nu_2}[l^2]^{\nu_3}[(l-p_3)^2]^{\nu_4}}
\end{split}\end{align}
for integer $\nu_i$ with $\nu_5,\nu_6,\nu_7\leq0$ and for $D=n-2\epsilon$, with $n$ even.
The space of functions defined by eq.~\eqref{eq:doubleEdgedGen} is spanned by four master
integrals, two of which are of the top topology, containing all four propagators, while the remaining two are the (massless) sunset integrals corresponding to pinching either of the two single-edge sides of the triangle.
These integrals can again be expressed in terms of Appell~$F_4$ functions, and we refer the reader to Section 5.3 in Ref.~\cite{Abreu:2021vhb} for details. The diagrammatic coaction takes the form
\begin{align}\nonumber
&\Delta\left[\begin{tikzpicture}[baseline={([yshift=-.5ex]current bounding box.center)}]
\coordinate (G1) at (0,0);
\coordinate (G2) at (1,0.57735026919);
\coordinate (G3) at (1,-0.57735026919);
\coordinate (H1) at (-1/3,0);
\coordinate [above right = 1/3 of G2](H2);
\coordinate [below right = 1/3 of G3](H3);
\coordinate (I1) at (1/2,0.28867513459);
\coordinate [above left=1/4 and 1/8 of I1](I11);
\coordinate [below right = 1/4 and 1/8 of I1](I12);
\coordinate (I2) at (1/2,-0.28867513459);
\coordinate [below left = 1/4 and 1/8 of I2](I21);
\coordinate [above right = 1/4 and 1/8 of I2](I22);
\coordinate (I3) at (1,0);
\coordinate (I31) at (3/4,0);
\coordinate (I32) at (5/4,0);
\coordinate [above left=1/3 of G2] (J1);
\draw (G1) -- (G2);
\draw (G1) -- (G3);
\draw (G2) to[out=-110,in=110] (G3);
\draw (G2) to[out=-70,in=70] (G3);
\draw (G1) [line width=0.75 mm] -- (H1);
\draw (G2) [line width=0.75 mm] -- (H2);
\draw (G3) [line width=0.75 mm] -- (H3);
\node at (H1) [left=0,scale=0.7] {$p_3$};
\node at (H2) [above right=0,scale=0.7] {$p_2$};
\node at (H3) [below right=0,scale=0.7] {$p_1$};
\node at (1/2,0.57735026919/2) [above left=0,scale=0.7] {\small$3$};
\node at (1/2,-0.57735026919/2) [below left=0,scale=0.7] {\small$4$};
\node at (0.7,0) [scale=0.7] {\small$1$};
\node at (1.3,0) [scale=0.7] {\small$2$};
\node at (J1) {\color{black!60!green}{\small$(1)$}};
\end{tikzpicture}\right]=
\begin{tikzpicture}[baseline={([yshift=-.5ex]current bounding box.center)}]
\coordinate (G1) at (0,0);
\coordinate (G2) at (1,0);
\coordinate (H1) at (-1/3,0);
\coordinate (H2) at (4/3,0);
\coordinate (I1) at (1/2,1/2);
\coordinate (I2) at (1/2,1/8);
\coordinate (I3) at (1/2,-1/2);
\coordinate (I4) at (1/2,-1/8);
\coordinate (J1) at (1,1/4);
\coordinate (J2) at (1,-1/4);
\coordinate (K1) at (1/2,-1/8);
\draw (G1) -- (G2);
\draw (G1) to[out=80,in=100] (G2);
\draw (G1) to[out=-80,in=-100] (G2);
\draw (G1) [line width=0.75 mm] -- (H1);
\draw (G2) [line width=0.75 mm]-- (H2);
\node at (K1) [above=0.7 mm of K1,scale=0.7] {\small$1$};
\node at (K1) [above=3.7 mm of K1,scale=0.7] {\small$4$};
\node at (K1) [below=1.3 mm of K1,scale=0.7] {\small$2$};
\node at (H1) [left=0,scale=0.7] {$p_1$};
\node at (H2) [right=0,scale=0.7] {$p_1$};
\end{tikzpicture}\otimes
\left(
\begin{tikzpicture}[baseline={([yshift=-.5ex]current bounding box.center)}]
\coordinate (G1) at (0,0);
\coordinate (G2) at (1,0.57735026919);
\coordinate (G3) at (1,-0.57735026919);
\coordinate (H1) at (-1/3,0);
\coordinate [above right = 1/3 of G2](H2);
\coordinate [below right = 1/3 of G3](H3);
\coordinate (I1) at (1/2,0.28867513459);
\coordinate [above left=1/4 and 1/8 of I1](I11);
\coordinate [below right = 1/4 and 1/8 of I1](I12);
\coordinate (I2) at (1/2,-0.28867513459);
\coordinate [below left = 1/4 and 1/8 of I2](I21);
\coordinate [above right = 1/4 and 1/8 of I2](I22);
\coordinate (I3) at (1,0);
\coordinate (I31) at (3/4,0);
\coordinate (I32) at (5/4,0);
\coordinate [above left=1/3 of G2] (J1);
\draw (G1) -- (G2);
\draw (G1) -- (G3);
\draw (G2) to[out=-110,in=110] (G3);
\draw (G2) to[out=-70,in=70] (G3);
\draw (G1) [line width=0.75 mm] -- (H1);
\draw (G2) [line width=0.75 mm] -- (H2);
\draw (G3) [line width=0.75 mm] -- (H3);
\draw (I21) [dashed,color=red,line width=0.5 mm] -- (I22);
\draw (I31) [dashed,color=red,line width=0.5 mm] -- (I3);
\draw (I32) [dashed,color=red,line width=0.5 mm] -- (I3);
\node at (H1) [left=0,scale=0.7] {$p_3$};
\node at (H2) [above right=0,scale=0.7] {$p_2$};
\node at (H3) [below right=0,scale=0.7] {$p_1$};
\node at (0.3,0.65/2) [scale=0.7] {\small$3$};
\node at (0.3,-0.65/2) [scale=0.7] {\small$4$};
\node at (0.75,0.2) [scale=0.7] {\small$1$};
\node at (1.25,0.2) [scale=0.7] {\small$2$};
\node at (J1) {\color{black!60!green}{\small$(1)$}};
\end{tikzpicture}
+\begin{tikzpicture}[baseline={([yshift=-.5ex]current bounding box.center)}]
\coordinate (G1) at (0,0);
\coordinate (G2) at (1,0.57735026919);
\coordinate (G3) at (1,-0.57735026919);
\coordinate (H1) at (-1/3,0);
\coordinate [above right = 1/3 of G2](H2);
\coordinate [below right = 1/3 of G3](H3);
\coordinate (I1) at (1/2,0.28867513459);
\coordinate [above left=1/4 and 1/8 of I1](I11);
\coordinate [below right = 1/4 and 1/8 of I1](I12);
\coordinate (I2) at (1/2,-0.28867513459);
\coordinate [below left = 1/4 and 1/8 of I2](I21);
\coordinate [above right = 1/4 and 1/8 of I2](I22);
\coordinate (I3) at (1,0);
\coordinate (I31) at (3/4,0);
\coordinate (I32) at (5/4,0);
\coordinate [above left=1/3 of G2] (J1);
\draw (G1) -- (G2);
\draw (G1) -- (G3);
\draw (G2) to[out=-110,in=110] (G3);
\draw (G2) to[out=-70,in=70] (G3);
\draw (G1) [line width=0.75 mm] -- (H1);
\draw (G2) [line width=0.75 mm] -- (H2);
\draw (G3) [line width=0.75 mm] -- (H3);
\draw (I11) [dashed,color=blue,line width=0.5 mm] -- (I12);
\draw (I21) [dashed,color=blue,line width=0.5 mm] -- (I22);
\draw (I31) [dashed,color=blue,line width=0.5 mm] -- (I3);
\draw (I32) [dashed,color=blue,line width=0.5 mm] -- (I3);
\node at (H1) [left=0,scale=0.7] {$p_3$};
\node at (H2) [above right=0,scale=0.7] {$p_2$};
\node at (H3) [below right=0,scale=0.7] {$p_1$};
\node at (0.3,0.65/2) [scale=0.7] {\small$3$};
\node at (0.3,-0.65/2) [scale=0.7] {\small$4$};
\node at (0.75,0.2) [scale=0.7] {\small$1$};
\node at (1.25,0.2) [scale=0.7] {\small$2$};
\node at (J1) {\color{black!60!green}{\small$(1)$}};
\end{tikzpicture}
\right)\\
&+\nonumber\!
\begin{tikzpicture}[baseline={([yshift=-.5ex]current bounding box.center)}]
\coordinate (G1) at (0,0);
\coordinate (G2) at (1,0);
\coordinate (H1) at (-1/3,0);
\coordinate (H2) at (4/3,0);
\coordinate (I1) at (1/2,1/2);
\coordinate (I2) at (1/2,1/8);
\coordinate (I3) at (1/2,-1/2);
\coordinate (I4) at (1/2,-1/8);
\coordinate (J1) at (1,1/4);
\coordinate (J2) at (1,-1/4);
\coordinate (K1) at (1/2,-1/8);
\draw (G1) -- (G2);
\draw (G1) to[out=80,in=100] (G2);
\draw (G1) to[out=-80,in=-100] (G2);
\draw (G1) [line width=0.75 mm] -- (H1);
\draw (G2) [line width=0.75 mm]-- (H2);
\node at (K1) [above=0.7 mm of K1,scale=0.7] {\small$1$};
\node at (K1) [above=3.7 mm of K1,scale=0.7] {\small$3$};
\node at (K1) [below=1.3 mm of K1,scale=0.7] {\small$2$};
\node at (H1) [left=0,scale=0.7] {$p_2$};
\node at (H2) [right=0,scale=0.7] {$p_2$};
\end{tikzpicture}\otimes
\left(\!\!
\begin{tikzpicture}[baseline={([yshift=-.5ex]current bounding box.center)}]
\coordinate (G1) at (0,0);
\coordinate (G2) at (1,0.57735026919);
\coordinate (G3) at (1,-0.57735026919);
\coordinate (H1) at (-1/3,0);
\coordinate [above right = 1/3 of G2](H2);
\coordinate [below right = 1/3 of G3](H3);
\coordinate (I1) at (1/2,0.28867513459);
\coordinate [above left=1/4 and 1/8 of I1](I11);
\coordinate [below right = 1/4 and 1/8 of I1](I12);
\coordinate (I2) at (1/2,-0.28867513459);
\coordinate [below left = 1/4 and 1/8 of I2](I21);
\coordinate [above right = 1/4 and 1/8 of I2](I22);
\coordinate (I3) at (1,0);
\coordinate (I31) at (3/4,0);
\coordinate (I32) at (5/4,0);
\coordinate [above left=1/3 of G2] (J1);
\draw (G1) -- (G2);
\draw (G1) -- (G3);
\draw (G2) to[out=-110,in=110] (G3);
\draw (G2) to[out=-70,in=70] (G3);
\draw (G1) [line width=0.75 mm] -- (H1);
\draw (G2) [line width=0.75 mm] -- (H2);
\draw (G3) [line width=0.75 mm] -- (H3);
\draw (I11) [dashed,color=red,line width=0.5 mm] -- (I12);
\draw (I31) [dashed,color=red,line width=0.5 mm] -- (I3);
\draw (I32) [dashed,color=red,line width=0.5 mm] -- (I3);
\node at (H1) [left=0,scale=0.7] {$p_3$};
\node at (H2) [above right=0,scale=0.7] {$p_2$};
\node at (H3) [below right=0,scale=0.7] {$p_1$};
\node at (0.3,0.65/2) [scale=0.7] {\small$3$};
\node at (0.3,-0.65/2) [scale=0.7] {\small$4$};
\node at (0.75,0.2) [scale=0.7] {\small$1$};
\node at (1.25,0.2) [scale=0.7] {\small$2$};
\node at (J1) {\color{black!60!green}{\small$(1)$}};
\end{tikzpicture}\!\!\!\!
+\begin{tikzpicture}[baseline={([yshift=-.5ex]current bounding box.center)}]
\coordinate (G1) at (0,0);
\coordinate (G2) at (1,0.57735026919);
\coordinate (G3) at (1,-0.57735026919);
\coordinate (H1) at (-1/3,0);
\coordinate [above right = 1/3 of G2](H2);
\coordinate [below right = 1/3 of G3](H3);
\coordinate (I1) at (1/2,0.28867513459);
\coordinate [above left=1/4 and 1/8 of I1](I11);
\coordinate [below right = 1/4 and 1/8 of I1](I12);
\coordinate (I2) at (1/2,-0.28867513459);
\coordinate [below left = 1/4 and 1/8 of I2](I21);
\coordinate [above right = 1/4 and 1/8 of I2](I22);
\coordinate (I3) at (1,0);
\coordinate (I31) at (3/4,0);
\coordinate (I32) at (5/4,0);
\coordinate [above left=1/3 of G2] (J1);
\draw (G1) -- (G2);
\draw (G1) -- (G3);
\draw (G2) to[out=-110,in=110] (G3);
\draw (G2) to[out=-70,in=70] (G3);
\draw (G1) [line width=0.75 mm] -- (H1);
\draw (G2) [line width=0.75 mm] -- (H2);
\draw (G3) [line width=0.75 mm] -- (H3);
\draw (I11) [dashed,color=blue,line width=0.5 mm] -- (I12);
\draw (I21) [dashed,color=blue,line width=0.5 mm] -- (I22);
\draw (I31) [dashed,color=blue,line width=0.5 mm] -- (I3);
\draw (I32) [dashed,color=blue,line width=0.5 mm] -- (I3);
\node at (H1) [left=0,scale=0.7] {$p_3$};
\node at (H2) [above right=0,scale=0.7] {$p_2$};
\node at (H3) [below right=0,scale=0.7] {$p_1$};
\node at (0.3,0.65/2) [scale=0.7] {\small$3$};
\node at (0.3,-0.65/2) [scale=0.7] {\small$4$};
\node at (0.75,0.2) [scale=0.7] {\small$1$};
\node at (1.25,0.2) [scale=0.7] {\small$2$};
\node at (J1) {\color{black!60!green}{\small$(1)$}};
\end{tikzpicture}
\right)\!
+\!
\begin{tikzpicture}[baseline={([yshift=-.5ex]current bounding box.center)}]
\coordinate (G1) at (0,0);
\coordinate (G2) at (1,0.57735026919);
\coordinate (G3) at (1,-0.57735026919);
\coordinate (H1) at (-1/3,0);
\coordinate [above right = 1/3 of G2](H2);
\coordinate [below right = 1/3 of G3](H3);
\coordinate (I1) at (1/2,0.28867513459);
\coordinate [above left=1/4 and 1/8 of I1](I11);
\coordinate [below right = 1/4 and 1/8 of I1](I12);
\coordinate (I2) at (1/2,-0.28867513459);
\coordinate [below left = 1/4 and 1/8 of I2](I21);
\coordinate [above right = 1/4 and 1/8 of I2](I22);
\coordinate (I3) at (1,0);
\coordinate (I31) at (3/4,0);
\coordinate (I32) at (5/4,0);
\coordinate [above left=1/3 of G2] (J1);
\draw (G1) -- (G2);
\draw (G1) -- (G3);
\draw (G2) to[out=-110,in=110] (G3);
\draw (G2) to[out=-70,in=70] (G3);
\draw (G1) [line width=0.75 mm] -- (H1);
\draw (G2) [line width=0.75 mm] -- (H2);
\draw (G3) [line width=0.75 mm] -- (H3);
\node at (H1) [left=0,scale=0.7] {$p_3$};
\node at (H2) [above right=0,scale=0.7] {$p_2$};
\node at (H3) [below right=0,scale=0.7] {$p_1$};
\node at (1/2,0.57735026919/2) [above left=0,scale=0.7] {\small$3$};
\node at (1/2,-0.57735026919/2) [below left=0,scale=0.7] {\small$4$};
\node at (0.7,0) [scale=0.7] {\small$1$};
\node at (1.3,0) [scale=0.7] {\small$2$};
\node at (J1) {\color{black!60!green}{\small$(1)$}};
\end{tikzpicture}\!\!\!\!\otimes\!
\begin{tikzpicture}[baseline={([yshift=-.5ex]current bounding box.center)}]
\coordinate (G1) at (0,0);
\coordinate (G2) at (1,0.57735026919);
\coordinate (G3) at (1,-0.57735026919);
\coordinate (H1) at (-1/3,0);
\coordinate [above right = 1/3 of G2](H2);
\coordinate [below right = 1/3 of G3](H3);
\coordinate (I1) at (1/2,0.28867513459);
\coordinate [above left=1/4 and 1/8 of I1](I11);
\coordinate [below right = 1/4 and 1/8 of I1](I12);
\coordinate (I2) at (1/2,-0.28867513459);
\coordinate [below left = 1/4 and 1/8 of I2](I21);
\coordinate [above right = 1/4 and 1/8 of I2](I22);
\coordinate (I3) at (1,0);
\coordinate (I31) at (3/4,0);
\coordinate (I32) at (5/4,0);
\coordinate [above left=1/3 of G2] (J1);
\draw (G1) -- (G2);
\draw (G1) -- (G3);
\draw (G2) to[out=-110,in=110] (G3);
\draw (G2) to[out=-70,in=70] (G3);
\draw (G1) [line width=0.75 mm] -- (H1);
\draw (G2) [line width=0.75 mm] -- (H2);
\draw (G3) [line width=0.75 mm] -- (H3);
\draw (I11) [dashed,color=black!60!green,line width=0.5 mm] -- (I12);
\draw (I21) [dashed,color=black!60!green,line width=0.5 mm] -- (I22);
\draw (I31) [dashed,color=black!60!green,line width=0.5 mm] -- (I3);
\draw (I32) [dashed,color=black!60!green,line width=0.5 mm] -- (I3);
\node at (H1) [left=0,scale=0.7] {$p_3$};
\node at (H2) [above right=0,scale=0.7] {$p_2$};
\node at (H3) [below right=0,scale=0.7] {$p_1$};
\node at (0.3,0.65/2) [scale=0.7] {\small$3$};
\node at (0.3,-0.65/2) [scale=0.7] {\small$4$};
\node at (0.75,0.2) [scale=0.7] {\small$1$};
\node at (1.25,0.2) [scale=0.7] {\small$2$};
\node at (J1) {\color{black!60!green}{\small$(1)$}};
\end{tikzpicture}
\\
&+\begin{tikzpicture}[baseline={([yshift=-.5ex]current bounding box.center)}]
\coordinate (G1) at (0,0);
\coordinate (G2) at (1,0.57735026919);
\coordinate (G3) at (1,-0.57735026919);
\coordinate (H1) at (-1/3,0);
\coordinate [above right = 1/3 of G2](H2);
\coordinate [below right = 1/3 of G3](H3);
\coordinate (I1) at (1/2,0.28867513459);
\coordinate [above left=1/4 and 1/8 of I1](I11);
\coordinate [below right = 1/4 and 1/8 of I1](I12);
\coordinate (I2) at (1/2,-0.28867513459);
\coordinate [below left = 1/4 and 1/8 of I2](I21);
\coordinate [above right = 1/4 and 1/8 of I2](I22);
\coordinate (I3) at (1,0);
\coordinate (I31) at (3/4,0);
\coordinate (I32) at (5/4,0);
\coordinate [above left=1/3 of G2] (J1);
\draw (G1) -- (G2);
\draw (G1) -- (G3);
\draw (G2) to[out=-110,in=110] (G3);
\draw (G2) to[out=-70,in=70] (G3);
\draw (G1) [line width=0.75 mm] -- (H1);
\draw (G2) [line width=0.75 mm] -- (H2);
\draw (G3) [line width=0.75 mm] -- (H3);
\node at (H1) [left=0,scale=0.7] {$p_3$};
\node at (H2) [above right=0,scale=0.7] {$p_2$};
\node at (H3) [below right=0,scale=0.7] {$p_1$};
\node at (1/2,0.57735026919/2) [above left=0,scale=0.7] {\small$3$};
\node at (1/2,-0.57735026919/2) [below left=0,scale=0.7] {\small$4$};
\node at (0.7,0) [scale=0.7] {\small$1$};
\node at (1.3,0) [scale=0.7] {\small$2$};
\node at (J1) {\color{blue}{\small$(2)$}};
\end{tikzpicture}\otimes\begin{tikzpicture}[baseline={([yshift=-.5ex]current bounding box.center)}]
\coordinate (G1) at (0,0);
\coordinate (G2) at (1,0.57735026919);
\coordinate (G3) at (1,-0.57735026919);
\coordinate (H1) at (-1/3,0);
\coordinate [above right = 1/3 of G2](H2);
\coordinate [below right = 1/3 of G3](H3);
\coordinate (I1) at (1/2,0.28867513459);
\coordinate [above left=1/4 and 1/8 of I1](I11);
\coordinate [below right = 1/4 and 1/8 of I1](I12);
\coordinate (I2) at (1/2,-0.28867513459);
\coordinate [below left = 1/4 and 1/8 of I2](I21);
\coordinate [above right = 1/4 and 1/8 of I2](I22);
\coordinate (I3) at (1,0);
\coordinate (I31) at (3/4,0);
\coordinate (I32) at (5/4,0);
\coordinate [above left=1/3 of G2] (J1);
\draw (G1) -- (G2);
\draw (G1) -- (G3);
\draw (G2) to[out=-110,in=110] (G3);
\draw (G2) to[out=-70,in=70] (G3);
\draw (G1) [line width=0.75 mm] -- (H1);
\draw (G2) [line width=0.75 mm] -- (H2);
\draw (G3) [line width=0.75 mm] -- (H3);
\draw (I11) [dashed,color=blue,line width=0.5 mm] -- (I12);
\draw (I21) [dashed,color=blue,line width=0.5 mm] -- (I22);
\draw (I31) [dashed,color=blue,line width=0.5 mm] -- (I3);
\draw (I32) [dashed,color=blue,line width=0.5 mm] -- (I3);
\node at (H1) [left=0,scale=0.7] {$p_3$};
\node at (H2) [above right=0,scale=0.7] {$p_2$};
\node at (H3) [below right=0,scale=0.7] {$p_1$};
\node at (0.3,0.65/2) [scale=0.7] {\small$3$};
\node at (0.3,-0.65/2) [scale=0.7] {\small$4$};
\node at (0.75,0.2) [scale=0.7] {\small$1$};
\node at (1.25,0.2) [scale=0.7] {\small$2$};
\node at (J1) {\color{black!60!green}{\small$(1)$}};
\end{tikzpicture},
\label{eq:diagCoactP1}
\end{align}
and similarly for the second master integral.
This example gives us an opportunity to discuss one of the key properties of the diagrammatic coaction, namely its behaviour in massless limits, which illustrates nicely how it captures the subtle properties of Feynman integrals.  The integral in eq.~(\ref{eq:doubleEdgedGen}) is infrared finite, as all external particles are off shell. However, the coaction (\ref{eq:diagCoactP1}) must be consistent for \emph{any} mass configuration, including in particular massless limits, in which infrared singularities arise. This property is of course realised at one loop. However, at two loop, additional complexity arises due to the fact that the number of master integrals varies as massless limits are taken. Specifically, while in the off-shell case eq.~(\ref{eq:diagCoactP1}) has four master integrals, two of which belong to the top topology, in the on-shell limit where say, $p_2^2=0$, there are only two master integrals, with only one at the top topology. This follows from the fact that in the massless limit one of the sunset integrals vanishes, and the two top-topology integrals become linearly dependent.
Recovering the coaction in this limit from eq.~(\ref{eq:diagCoactP1}) requires intricate relations between cuts, which must be valid to all orders in $\epsilon$. This is indeed realised, and one obtains the following diagrammatic coaction for $p_2^2=0$,
\begin{align}\label{eq;diagCoaction2Asy}
&\Delta\left[\begin{tikzpicture}[baseline={([yshift=-.5ex]current bounding box.center)}]
\coordinate (G1) at (0,0);
\coordinate (G2) at (1,0.57735026919);
\coordinate (G3) at (1,-0.57735026919);
\coordinate (H1) at (-1/3,0);
\coordinate [above right = 1/3 of G2](H2);
\coordinate [below right = 1/3 of G3](H3);
\coordinate (I1) at (1/2,0.28867513459);
\coordinate [above left=1/4 and 1/8 of I1](I11);
\coordinate [below right = 1/4 and 1/8 of I1](I12);
\coordinate (I2) at (1/2,-0.28867513459);
\coordinate [below left = 1/4 and 1/8 of I2](I21);
\coordinate [above right = 1/4 and 1/8 of I2](I22);
\coordinate (I3) at (1,0);
\coordinate (I31) at (3/4,0);
\coordinate (I32) at (5/4,0);
\coordinate [above left=1/4 of G2] (J1);
\draw (G1) -- (G2);
\draw (G1) -- (G3);
\draw (G2) to[out=-110,in=110] (G3);
\draw (G2) to[out=-70,in=70] (G3);
\draw (G1) [line width=0.75 mm] -- (H1);
\draw (G2) -- (H2);
\draw (G3) [line width=0.75 mm]-- (H3);
\node at (H1) [left=0,scale=0.7] {$p_3$};
\node at (H2) [above right=0,scale=0.7] {$p_2$};
\node at (H3) [below right=0,scale=0.7] {$p_1$};
\node at (1/2,0.57735026919/2) [above left=0,scale=0.7] {\small$3$};
\node at (1/2,-0.57735026919/2) [below left=0,scale=0.7] {\small$4$};
\node at (0.7,0) [scale=0.7] {\small$1$};
\node at (1.3,0) [scale=0.7] {\small$2$};
\end{tikzpicture}\right]
=\begin{tikzpicture}[baseline={([yshift=-.5ex]current bounding box.center)}]
\coordinate (G1) at (0,0);
\coordinate (G2) at (1,0.57735026919);
\coordinate (G3) at (1,-0.57735026919);
\coordinate (H1) at (-1/3,0);
\coordinate [above right = 1/3 of G2](H2);
\coordinate [below right = 1/3 of G3](H3);
\coordinate (I1) at (1/2,0.28867513459);
\coordinate [above left=1/4 and 1/8 of I1](I11);
\coordinate [below right = 1/4 and 1/8 of I1](I12);
\coordinate (I2) at (1/2,-0.28867513459);
\coordinate [below left = 1/4 and 1/8 of I2](I21);
\coordinate [above right = 1/4 and 1/8 of I2](I22);
\coordinate (I3) at (1,0);
\coordinate (I31) at (3/4,0);
\coordinate (I32) at (5/4,0);
\coordinate [above left=1/4 of G2] (J1);
\draw (G1) -- (G2);
\draw (G1) -- (G3);
\draw (G2) to[out=-110,in=110] (G3);
\draw (G2) to[out=-70,in=70] (G3);
\draw (G1) [line width=0.75 mm] -- (H1);
\draw (G2) -- (H2);
\draw (G3) [line width=0.75 mm]-- (H3);
\node at (H1) [left=0,scale=0.7] {$p_3$};
\node at (H2) [above right=0,scale=0.7] {$p_2$};
\node at (H3) [below right=0,scale=0.7] {$p_1$};
\node at (1/2,0.57735026919/2) [above left=0,scale=0.7] {\small$3$};
\node at (1/2,-0.57735026919/2) [below left=0,scale=0.7] {\small$4$};
\node at (0.7,0) [scale=0.7] {\small$1$};
\node at (1.3,0) [scale=0.7] {\small$2$};
\end{tikzpicture}\!\!
\otimes\begin{tikzpicture}[baseline={([yshift=-.5ex]current bounding box.center)}]
\coordinate (G1) at (0,0);
\coordinate (G2) at (1,0.57735026919);
\coordinate (G3) at (1,-0.57735026919);
\coordinate (H1) at (-1/3,0);
\coordinate [above right = 1/3 of G2](H2);
\coordinate [below right = 1/3 of G3](H3);
\coordinate (I1) at (1/2,0.28867513459);
\coordinate [above left=1/4 and 1/8 of I1](I11);
\coordinate [below right = 1/4 and 1/8 of I1](I12);
\coordinate (I2) at (1/2,-0.28867513459);
\coordinate [below left = 1/4 and 1/8 of I2](I21);
\coordinate [above right = 1/4 and 1/8 of I2](I22);
\coordinate (I3) at (1,0);
\coordinate (I31) at (3/4,0);
\coordinate (I32) at (5/4,0);
\coordinate [above left=1/4 of G2] (J1);
\draw (G1) -- (G2);
\draw (G1) -- (G3);
\draw (G2) to[out=-110,in=110] (G3);
\draw (G2) to[out=-70,in=70] (G3);
\draw (G1) [line width=0.75 mm] -- (H1);
\draw (G2) -- (H2);
\draw (G3) [line width=0.75 mm]-- (H3);
\draw (I11) [dashed,color=violet,line width=0.5 mm] -- (I12);
\draw (I21) [dashed,color=violet,line width=0.5 mm] -- (I22);
\draw (I31) [dashed,color=violet,line width=0.5 mm] -- (I3);
\draw (I32) [dashed,color=violet,line width=0.5 mm] -- (I3);
\node at (H1) [left=0,scale=0.7] {$p_3$};
\node at (H2) [above right=0,scale=0.7] {$p_2$};
\node at (H3) [below right=0,scale=0.7] {$p_1$};
\node at (0.3,0.65/2) [scale=0.7] {\small$3$};
\node at (0.3,-0.65/2) [scale=0.7] {\small$4$};
\node at (0.75,0.2) [scale=0.7] {\small$1$};
\node at (1.25,0.2) [scale=0.7] {\small$2$};
\end{tikzpicture}\!\!+\begin{tikzpicture}[baseline={([yshift=-.5ex]current bounding box.center)}]
\coordinate (G1) at (0,0);
\coordinate (G2) at (1,0);
\coordinate (H1) at (-1/3,0);
\coordinate (H2) at (4/3,0);
\coordinate (I1) at (1/2,1/2);
\coordinate (I2) at (1/2,1/8);
\coordinate (I3) at (1/2,-1/2);
\coordinate (I4) at (1/2,-1/8);
\coordinate (J1) at (1,1/4);
\coordinate (J2) at (1,-1/4);
\coordinate (K1) at (1/2,-1/8);
\draw (G1) -- (G2);
\draw (G1) to[out=80,in=100] (G2);
\draw (G1) to[out=-80,in=-100] (G2);
\draw (G1) [line width=0.75 mm] -- (H1);
\draw (G2) [line width=0.75 mm]-- (H2);
\node at (K1) [above=0.7 mm of K1,scale=0.7] {\small$1$};
\node at (K1) [above=3.7 mm of K1,scale=0.7] {\small$4$};
\node at (K1) [below=1.3 mm of K1,scale=0.7] {\small$2$};
\node at (H1) [left=0,scale=0.7] {$p_1$};
\node at (H2) [right=0,scale=0.7] {$p_1$};
\end{tikzpicture}\otimes\begin{tikzpicture}[baseline={([yshift=-.5ex]current bounding box.center)}]
\coordinate (G1) at (0,0);
\coordinate (G2) at (1,0.57735026919);
\coordinate (G3) at (1,-0.57735026919);
\coordinate (H1) at (-1/3,0);
\coordinate [above right = 1/3 of G2](H2);
\coordinate [below right = 1/3 of G3](H3);
\coordinate (I1) at (1/2,0.28867513459);
\coordinate [above left=1/4 and 1/8 of I1](I11);
\coordinate [below right = 1/4 and 1/8 of I1](I12);
\coordinate (I2) at (1/2,-0.28867513459);
\coordinate [below left = 1/4 and 1/8 of I2](I21);
\coordinate [above right = 1/4 and 1/8 of I2](I22);
\coordinate (I3) at (1,0);
\coordinate (I31) at (3/4,0);
\coordinate (I32) at (5/4,0);
\coordinate [above left=1/4 of G2] (J1);
\draw (G1) -- (G2);
\draw (G1) -- (G3);
\draw (G2) to[out=-110,in=110] (G3);
\draw (G2) to[out=-70,in=70] (G3);
\draw (G1) [line width=0.75 mm] -- (H1);
\draw (G2) -- (H2);
\draw (G3) [line width=0.75 mm]-- (H3);
\draw (I21) [dashed,color=violet,line width=0.5 mm] -- (I22);
\draw (I31) [dashed,color=violet,line width=0.5 mm] -- (I3);
\draw (I32) [dashed,color=violet,line width=0.5 mm] -- (I3);
\node at (H1) [left=0,scale=0.7] {$p_3$};
\node at (H2) [above right=0,scale=0.7] {$p_2$};
\node at (H3) [below right=0,scale=0.7] {$p_1$};
\node at (0.3,0.65/2) [scale=0.7] {\small$3$};
\node at (0.3,-0.65/2) [scale=0.7] {\small$4$};
\node at (0.75,0.2) [scale=0.7] {\small$1$};
\node at (1.25,0.2) [scale=0.7] {\small$2$};
\end{tikzpicture}\!\!,
\end{align}
where each entry evaluates to Gauss hypergeometric functions, as can be seen either by a direct calculation, or as a limit of eq.~(\ref{eq:diagCoactP1}). 

Having seen a few examples, let us now summarise some key features of the diagrammatic coaction beyond one loop, which we write as~\cite{Abreu:2021vhb} 
\begin{equation}\label{eq:coac_L_loop}
	\Delta\left(\int_{{\color[rgb]{0.1,0.1,1.0} \Gamma_\emptyset}}{\color[rgb]{1.0,0.1,0.1}\omega_G^{(k)}}\right)=
	\sum_{C\in M_G}\sum_i
	%{i=1}^{\displaystyle{\kappa_{G_C}}}
	\int_{{\color[rgb]{0.1,0.1,1.0} \Gamma_\emptyset}}\omega_{G_C}^{(i)}\otimes
	\int_{\gamma_C^{(i)}}{\color[rgb]{1.0,0.1,0.1}\omega_G^{(k)}}\qquad \text{with}\qquad
\gamma_C^{(k)}=\sum_{{\substack{X\in M_G\\ C\subseteq X}}}
	\sum_{i}
%	^{\displaystyle{\kappa_{G_X}}}
\alpha_X^{(k,i)}\Gamma_X^{(i)}\,,
\end{equation}
where~$i$ indexes the elements of the basis forms~$\omega_{G_C}^{(i)}$ for a given $C$, as well as their dual contours~$\gamma_C^{(i)}$.
This formula reduces to eq.~\eqref{Diagr_coaction_one_loop} for one-loop integrals: in that case~$M_G$ corresponds to all non-empty subsets of edges of the graph $G$ and there is just a single value of $i$ for every~$C$.
The dual contour to $\omega^{(k)}_{G_C}$ is a linear combination of contours
that encircle all the poles of the propagators in~$C$ or more (but not fewer).
The coefficients $\alpha_X^{(k,i)}$ can in general depend on the same variables as the Feynman 
integral.

We stress that while the one-loop coaction~\eqref{Diagr_coaction_one_loop}
is fully explicit (the bases have been fixed and all integrals have been explicitly defined for a generic mass configuration and any number of legs, see Refs.~\cite{Abreu:2017enx,Abreu:2017mtm}), the $L$-loop generalisation
\eqref{eq:coac_L_loop} is not.
Beyond one loop the set of master integrals and their dual contours needs to be identified on a case-by-case basis. 
Nevertheless, some properties of the coaction are understood in 
general. In particular, all left entries are master integrals with $L$ loops of the given topology, and  
all propagators that feature on a given left entry are cut on the corresponding right one. 

While multi-loop homology relations, which dictate the details of the coaction, are not known in general, certain relations follow directly from the one-loop case.  Ref.~\cite{Abreu:2021vhb}  defines~\emph{genuine L-loop cuts} as cuts which leave no loop uncut. Because the one-loop relation of eq.~(\ref{eq:PCI}) may be applied to any loop of a multi-loop integral, we can always establish a basis of cuts solely in terms of genuine $L$-loop cuts, and use it to define the right entries in the coaction.

\section{Conclusions}

We constructed a coaction on integrals (\ref{int_coaction}) based on pairing between master integrands and master contours. 
This coaction naturally applies to generalised hypergeometric functions including all Appell functions, where $\epsilon$ dependence is introduced in the framework of twisted (co)-homology. This coaction reproduces the coaction on MPLs upon expansion. 
It also translates into a coaction of dimensionally-regularized Feynman integrals for any one-loop diagram, with any mass configuration. Right entries are master integrals while left entries are cuts. 
Relations between cuts (homology) are essential to establish its precise form.

We conjecture that the diagrammatic coaction extends to the multi-loop case according to eq.~(\ref{eq:coac_L_loop}), identifying key features:
the left entries are the master integrals of the given topology, while the right entries are cuts, where all propagators that feature in a given left entry must be cut on the corresponding right entry. In contrast with one loop, the bases need to be set for each topology. While there is no complete theory of the relations between contours (cuts) for multi-loop integrals, certain features can be deduced from the one-loop case in a loop-by-loop analysis. Specifically, the basis of cuts can be chosen in terms of genuine $L$-loop cuts, where every loop features at least one cut propagator.  The diagrammatic coaction is consistent with massless limits, and it encodes discontinuities and differential equations of Feynman integrals. The analysis of more complex topologies ~\cite{Gardi:2021qov} and elliptic cases is under way. 

Besides the fundamental nature of this study in understanding the algebraic and analytic structure of Feynman integrals, there are several important applications. Some applications of the coaction in computing integrals have been initiated in Ref.~\cite{Abreu:2017mtm} in the context of multi-leg one-loop integrals, including a determination of the differential equation in terms of cuts (see Figure~\ref{fig:Pentagon_DE}  above) as well as an iterative expression for the symbol of such integrals.  
Another set of applications is the use of the duality between master integrands and cut contours to project the integrand of an amplitude into a set of master integrals, by-passing the need to solve the integration-by-parts system. Using intersection theory to this end has been an active research direction recently~\cite{Mizera:2017rqa,Mizera:2019gea,Mastrolia:2018uzb,Caron-Huot:2021xqj,Caron-Huot:2021iev,Frellesvig:2019uqt,Weinzierl:2020xyy,Chen:2020uyk,Frellesvig:2020qot,Weinzierl:2020gda,Frellesvig:2021vem,Mandal:2022vok,Chen:2022lzr}, but the precise connection to the diagrammatic coaction has not yet been studied.

\bibliographystyle{JHEP}
\bibliography{bibMain_LL}

\end{document}